\documentclass[aps,prb,reprint,floatfix,superscriptaddress]{revtex4-2}

\usepackage[english]{babel} % Language hyphenation and typographical rules
\usepackage{hyperref}
\usepackage{amssymb,amsmath}
\usepackage{graphicx}
\usepackage{float}
\usepackage{url}
\usepackage{multirow}
\usepackage{chemformula}
\usepackage{tabularx}
\usepackage{placeins}
\usepackage{mathtools}

\usepackage[utf8]{inputenc}

\newlength{\smallpic}
\begin{document}
\newcommand{\dprime}{{\prime \prime}}

\title{Predicting Néel temperatures in helimagnetic materials}
% \title{Predicting the Néel temperatures in general helimagnetic materials: a comparison between mean field theory, random phase approximation, renormalized spin wave theory and classical Monte Carlo simulations}
\author{Varun Rajeev Pavizhakumari}
\affiliation{CAMD, Computational Atomic-Scale Materials Design, Department of Physics, Technical University of Denmark, 2800 Kgs. Lyngby Denmark}
\author{Thomas Olsen}
\email{tolsen@fysik.dtu.dk}
\affiliation{CAMD, Computational Atomic-Scale Materials Design, Department of Physics, Technical University of Denmark, 2800 Kgs. Lyngby Denmark}

\begin{abstract}
The critical temperature for magnetic order comprises a crucial property of any magnetic material and ranges from a few Kelvin in certain antiferromagnets to 1400 K in ferromagnetic Co. However, the prediction of critical temperatures based on, for example, a spin wave dispersion is in general non-trivial. For ferromagnets and simple collinear antiferromagnets, estimates may be obtained from the Heisenberg model using either renormalized spin wave theory or the Green's function random phase approximation (RPA), but a systematic assessment of the accuracy of such approaches seems to be lacking in the literature. In this work, we propose generalizations of both renormalized spin wave theory and RPA to calculate the critical temperatures of single-$Q$ helimagnetic ground states, which include ferromagnets and antiferromagnets as special cases. We compare the methods to classical Monte Carlo simulations and Mean field theory, using experimental exchange parameters for a wide range of materials; MnO and NiO (single site Néel ground states), MnF$_2$ (altermagnet), Cr$_2$O$_3$ and Fe$_2$O$_3$ (two site Néel states) and  Ba$_3$NbFe$_3$Si$_2$O$_{14}$ (incommensurate helimagnet). In all cases, we observe that predictions from RPA are in excellent agreement with experimental values and RPA  thus constitutes a rather reliable all-purpose method for calculating critical temperatures.
\end{abstract}

\maketitle
\section{Introduction}

Magnetic order can manifest itself in many different varieties and is typically determined by the nature of interactions between localized magnetic moments. Apart from conventional scenarios of ferromagnetic and collinear antiferromagnetic order, competing interactions and geometric frustration may lead to non-collinear magnetic order. In this respect, the helically ordered states (single-$Q$ planar spin spirals) comprise a rather general class of ground states that minimizes the {\it classical} energy of the isotropic Heisenberg model and includes ferromagnetic and antiferromagnetic order as special cases. Due to broken symmetries, such structures can have a variety of physical properties such as type II multiferroicity \cite{khomskii2009classifying} and odd-parity altermagnetism \cite{PhysRevX.12.040501}. Collective excitations in the ordered state form spin waves or magnons and they exist in the low end of the energy spectrum, making them advantageous for low power consumption applications. 
However, the technological relevance of a magnetic material crucially depends on the thermal stability of spin waves at operating temperature (typically room temperature), and consequently there is considerable interest in methods for accurately predicting the critical temperature for magnetic order. Although well-known predictive schemes exist for the special cases of ferromagnets and antiferromagnets, there does not seem to be a general framework that accurately deals with arbitrary helimagnetic ground states of the Heisenberg model. 
The focus of the present work will be to introduce such a framework and to benchmark it against experimentally known Néel temperatures of materials that allow for a reliable Heisenberg model description.

The theory of phase transitions in magnetism can be traced back to Pierre Curie's works on ferromagnetism \cite{Curie1894}. Following this, Weiss' mean (molecular) field theory provided an explanation for spontaneous magnetic order in ferromagnetic materials \cite{Weiss1907}, but the assumption of a uniform molecular field cannot deal with antiferromagnetic order. This was addressed by Louis Néel, who introduced the concept of a local exchange field that alternates in sign between sites such that anti-alignment between neighboring moments is favored in the ground state (the Néel state) \cite{Neel1936,Neel1971}. Finally, the concept of a local molecular field is easily generalized to helimagnetic ground states, which allows for a mean field treatment of critical temperatures, susceptibilities, and heat capacities in general helimagnets \cite{Johnston2015}.

While the simplicity of the mean field treatment is appealing, the quantitative agreement with experiments is often questionable due to the complete neglect of correlation effects. A more accurate treatment can be obtained by representing the spin operators in the Heisenberg model in terms of bosonic operators - a procedure known as the Holstein-Primakoff (HP) transformation \cite{hp1940}. Truncating the model at second order in the bosonic operators renders the model exactly solvable and the fundamental excitations are referred to as magnons. This procedure is rather straightforward for ferromagnets, where the ground state coincides with the classical spin configuration that minimizes the energy. 
However, the quantum mechanical ground state of antiferromagnets is not known exactly (except for one-dimensional spin-1/2 systems) and the HP transformation has to rely on the {\it classical} ground state energy of the Heisenberg model. Diagonalizing the bosonic Hamitonian that results from a second order truncation of the operators then yields a new ground state with an energy below the classical ground state and a new set of bosonic operators that are related to the original ones by a Bogoliubov transformation \cite{Anderson1952,Marshall1955,Watabe1995}. While the HP transformation has been carried out for various specific cases of antiferromagnetic order \cite{Anderson1952,Marshall1955,Watabe1995}, a complete generalization to any helimagnetic order was derived by Toth and Lake \cite{Toth2015} using a rotated reference frame. This method includes ferromagnetic and antiferromagnetic order as special cases and is implemented in the spinW code \cite{Toth2015}.
In order to include thermal effects, in the model one has to go beyond the second order approximation and include higher order terms in bosonic operators, which are naturally interpreted as magnon interactions. In general, such terms render the model intractable, but one may include fourth order terms in a mean field treatment, which yields a non-interacting bosonic Hamiltonian that is to be solved self-consistently for the thermally renormalized magnons. This procedure is well known for ferromagnets \cite{yosida2010theory, beyondrpa} and yields the critical temperature as the point where thermal occupation of magnons equals the ground state magnetization, but a general framework for helimagnetic ground state order has (to the best of our knowledge) not been presented prior to this work. 

An alternative to the renormalized HP theory described above is the Green's function method introduced by Bogoliubov and Tyablikov \cite{tyablikov1959}. In the present work we will refer to this as the random phase approximation (RPA) although the renormalized HP approach is occasionally also referred to as RPA \cite{yosida2010theory}. The method is based on the equation of motion for the magnetic susceptibility (Green's function), which is solved by a mean field approximation for higher order correlation functions. The poles of the resulting Green's functions depend on temperature and may be interpreted as renormalized magnon energies. The Green's function method was originally developed for ferromagnetic order \cite{tyablikov1959,beyondrpa,tahir1962} and later simplified by Callen \cite{Callen1963}. The approach was rapidly generalized to collinear antiferromagnetic order \cite{Lee1967} and single site helimagnetic order \cite{Turek2003}, but a comprehensive general framework for helimagnetic order still seems to be lacking in the literature.

The fact that the number of magnetic sites scales with the size of the magnetic unit cell comprises a major challenge in the treatment of spinwave excitations in helical ground states. Moreover, for an incommensurate spin spiral it is impossible to represent the magnetic order exactly in a magnetic super cell and any attempt of representing the ground state by an approximate rational ordering vector will typically lead to a very large magnetic super cell. However, this can be tackled by representing the problem in a rotating frame in which all moments are aligned along the same (local) axis. This greatly simplifies the problem and allows one to calculate the magnon energies with reference to the crystallographic unit cell.  
Such an approach was introduced in Ref. \onlinecite{Toth2015} in the non-interacting magnon limit for the HP method. In the present work, we will include the magnon interactions in the HP model and obtain the thermally renormalized magnon energies for spin spiral ground states. Similarly, we introduce a generalized Green's function method for spin spirals which only refers to the crystallographic unit cell and both approaches will allow us to calculate the Néel temperatures for any single-$Q$ ordered compound. In order to benchmark the performance of these methods, we consider a diverse range of antiferromagnetic/helimagnetic materials with experimentally known exchange interactions and Néel temperatures. We take MnO and NiO as representatives of compounds with one magnetic site in the crystallographic unit cell. Cr$_2$O$_3$ and Fe$_2$O$_3$ are chosen as representatives of isostructural compounds with two magnetic sites but different ordering vectors. We also consider MnF$_2$ which exhibits an altermagnetic ground state \cite{MnF2-Altermagnet} and show that the (small) altermagnetic magnon splitting has minimal effect on the calculated critical temperature. In addition, we apply the method to Ba$_3$NbFe$_3$Si$_2$O$_{14}$, which has a complex magnetic structure with three magnetic sites in the crystallographic unit cell and an incommensurate ordering vector. We compare the calculated critical temperatures with results obtained from classical Monte Carlo simulations and mean field theory as well as the experimental values.

This paper is organized as follows. In Sec. \ref{sec:HelicalGS}, we briefly discuss the Heisenberg model and single-$Q$ spiral ground states. We then introduce the methodologies for calculating thermal properties of spin spiral structures using renormalized HP bosononization (Sec. \ref{sec:HP}), the Green's function method Sec. \ref{sec:Gfn}) and the Mean Field approximation (Sec. \ref{sec:MFA}). In Sec. \ref{sec:results}, we benchmark the methods against the experimental results for a range of real materials.

\section{Theory}
\subsection{Spiral ground states}\label{sec:HelicalGS}
Our starting point is the generalized Heisenberg Hamiltonian
\begin{equation}\label{eq:heisenberg-model}
    \mathcal{H} = -\frac{1}{2}\sum_{abij} \mathbf{S}^{T}_{ai} \textrm{J}_{abij} \mathbf{S}_{bj},
\end{equation}
where $\mathbf{S}_{ai}$ is the spin operator for site $a$ in unit cell $i$ and $\textrm{J}_{abij}$ is the exchange tensor that couples the spins at sites $a,i$ and  $b,j$. Single-ion anisotropy terms are included through the onsite components $\textrm{J}_{aaii}$ of the exchange tensor. 
In order to calculate the thermal properties of the model \eqref{eq:heisenberg-model} one needs the magnetic excitation spectrum, which requires one to determine the magnetic ground state. Except for simple ferromagnets, the ground state of the model is non-trivial and the starting point will thus be the spin configuration that minimizes the {\it classical} energy. 
For an isotropic model on a Bravais lattice, the minimizing spin configuration can be determined using the Luttinger-Tisza (LT) method \cite{Luttinger1946} and the result is known to be a coplanar spin spiral characterized by a magnetic ordering vector $\mathbf{Q}$ \cite{yosida2010theory}. The result can be generalized to include antisymmetric exchange, which then determines the orientation of the spiral plane \cite{PhysRevB.94.024403}. 
The LT method has also been generalized to non-Bravais lattices \cite{Lyons1960, Schmidt2022} and under rather general circumstances (but not always) the classical ground state for an isotropic model remains a planar spin spiral given by
\begin{equation}\label{eq:spin-spiral-gs}
    \mathbf{S}_{ai} = S_a 
    \begin{bmatrix}
        0\\ \textrm{sin}(\mathbf{Q}\cdot \mathbf{r}_i + \phi_a)\\ \textrm{cos}(\mathbf{Q}\cdot \mathbf{r}_i + \phi_a) 
    \end{bmatrix}, 
\end{equation}
where $\mathbf{Q}$ is the magnetic ordering vector,  $\phi_a$ is the phase angle of the spin at site $a$, $\mathbf{r}_i$ is a lattice vector for unit cell $i$, and $S_a$ is the maximal eigenvalue of $S_a^z$ (the magnitude of the classical spin vector). We note that the assumption of an isotropic model implies that any orientation of the spiral plane yields the same ground state energy and we have chosen the $yz$-plane here to be specific. Such a magnetic structure is often referred as a single-$Q$ spin spiral and is shown to be the ground state in appendix \ref{app:Classical ground state}. The discussion in the remainder of this work will focus solely on single-$Q$ spin spirals.

When considering excitations of spiral ground states it will be convenient to transform the system to a local coordinate system where all spin fluctuations are perpendicular to the local $z$-axis. Here we reiterate the approach of Ref. \onlinecite{Toth2015} regarding this transformation, which will be needed for reference below. We can write the spin operators in terms of a cell-dependent rotation matrix $\textrm{R}_i$ such that 
\begin{align}
     \mathbf{S}_{ai} &= \textrm{R}_i \mathbf{S}'_{ai} \label{eq:rot-uc},\\
     \textrm{R}_i &= \textrm{R}_\mathbf{\hat{n}}(\mathbf{Q}\cdot \mathbf{r}_i),
\end{align}
where $\mathbf{\hat{n}}$ is the global rotation axis. The classical analogues of the $\mathbf{S}'_{ai}$ operators thus exhibit a ground state configuration that is independent of the unit cell index $i$. We then define another rotation:
\begin{align}
    \mathbf{S}^\prime_{ai} &= \textrm{R}^\prime_a \mathbf{S}^\dprime_{ai}, \label{eq:rot-site} 
\end{align}
such that $ \textrm{R}^\prime_a$ generates the classical spin configuration within a single unit cell from a set of aligned spins $\mathbf{S}^\dprime_{ai}$-chosen along the $z$-direction here. The classical analogue of $\mathbf{S}^\dprime_{ai}$ describes a locally ferromagnetic ground state configuration where all spins are aligned along the $z$-direction. 
It is convenient to introduce two vectors $\mathbf{u}_a$ and $\mathbf{v}_a$ defined by the columns of the site rotation matrices in Eq. \eqref{eq:rot-site}:
\begin{align}
    [\mathbf{u}_a]^\mu &= \big[{\textrm{R}^\prime_a}\big]^{\mu x} + i\big[{\textrm{R}^\prime_a}\big]^{\mu y} \label{eq:u_a},\\
    [\mathbf{v}_a]^\mu &= \big[{\textrm{R}^\prime_a}\big]^{\mu z}\label{eq:v_a}.
\end{align}
In terms of these the Heisenberg model \eqref{eq:heisenberg-model} can be written as
\begin{align}
    \mathcal{H} &= -\frac{1}{2}\sum_{abij} \mathbf{S}^{\prime T}_{ai}  \textrm{R}^{T}_{i}\textrm{J}_{abij} \textrm{R}_{j}\mathbf{S}^{\prime}_{bj}\label{eq:rot-heisenberg-model0},
\end{align}
and we may use Eqs. \eqref{eq:u_a}-\eqref{eq:v_a} to express Eq. \eqref{eq:rot-site} as
\begin{align}\label{eq:Sprime-uv}
    \mathbf{S}^{\prime}_{ai} = \mathbf{u}_a \frac{S^{\dprime +}_{ai}}{2} + \mathbf{u}^*_a \frac{S^{\dprime -}_{ai}}{2} + \mathbf{v}_a S^{\dprime z}_{ai},
\end{align}
with
\begin{align}
    S^{\dprime +}_{ai} &= S^{\dprime x}_{ai} + iS^{\dprime y}_{ai},\\
    S^{\dprime -}_{ai} &= S^{\dprime x}_{ai} - iS^{\dprime y}_{ai}.
\end{align}
This allow us to retain the crystallographic unit cell and the complexity implied by the spiral order will be transformed to a locally rotated exchange tensor. In particular, we may calculate thermal properties in much simpler terms compared to working with the full magnetic unit cell. Below we will address the problem using three different approximate methods: 
Holstein-Primakoff bosonization, Green's function RPA and the mean field approximation.

\subsection{Holstein-Primakoff bosonization}\label{sec:HP}
In the Holstein-Primakoff bosonization, the spin operators are written in terms of bosonic creation and annihilation operators such that:
\begin{align}
    S^{\dprime +}_{ai} &= \sqrt{2S_a} \sqrt{1-\frac{a_{ai}^\dag a_{ai}}{2S_a}} a_{ai}  \label{eq:hpS+},\\
    S^{\dprime -}_{ai} &= \sqrt{2S_a} a^\dagger_{ai} \sqrt{1-\frac{a_{ai}^\dag a_{ai}}{2S_a}\label{eq:hpS-}},\\
    S^{\dprime z}_{ai} &= S_a - a^\dagger_{ai} a_{ai} \label{eq:hpSz}.
\end{align}
These operators, obey the usual commutation relations $[a_{ai},a^\dagger_{bj}]=\delta_{ij}\delta_{ab}$ and  $[a_{ai},a_{bj}]=0$, which guarantee that the spin commutator relations are satisfied. It should be noted that $S^{\dprime +}_{ai}$ and $S^{\dprime -}_{ai}$ always raise and lower the spin along the {\it local} $z$-axis. The $a_{ai}$ operators thus have the property that they annihilate the ``spiral Néel state", which is defined as the state that satisfies $\langle S^{\dprime z}_{ai}\rangle=S_a$. This state coincides with the quantum mechanical ground state for a ferromagnet, but in general it should be regarded as a classical ground state spin configuration, which may not be an eigenstate of the Hamiltonian. Nevertheless, it will serve as the reference state on which the bosonic operators  \eqref{eq:hpS+}-\eqref{eq:hpSz} have a well-defined action. 
We apply the HP transformation \eqref{eq:hpS+}-\eqref{eq:hpSz} to Eq. \eqref{eq:Sprime-uv} and expand the square roots to third order in bosonic operators:
\begin{align}\label{eq:Sprime-uv-hp}
    \mathbf{S}^\prime_{ai}  &\approx \mathbf{u}^{*T}_a \Big(\sqrt{\frac{S_a}{2}}a_{ai} - \frac{a^\dagger_{ai} a_{ai} a_{ai}}{4\sqrt{2S_a}} \Big)   \notag \\& + \mathbf{u}^T_a \Big(\sqrt{\frac{S_a}{2}}a^\dagger_{ai} - \frac{a^\dagger_{ai} a^\dagger_{ai} a_{ai}}{4\sqrt{2S_a}} \Big)  \notag \\& + \mathbf{v}^T_a(S_a - a^\dagger_{ai} a_{ai} ).
\end{align}
Inserting this into the Hamiltonian \eqref{eq:rot-heisenberg-model0}, yields a model, which we truncate at fourth order in bosonic operator products. The first order terms do not contribute to the magnon energies as they can be removed by a linear transformation. Upon truncating the expansion at second order in bosonic operators, the Hamiltonian is readily diagonalized and the resulting eigenstates are interpreted as magnons. 
This approximation comprises the non-interacting magnon limit, which is accurate at low temperatures, but fails at describing thermal properties at elevated temperatures. To improve upon this, one can include terms up to fourth order in bosonic operators, which may be interpreted as magnon interactions. These higher order terms are then treated in a Hartree-Fock type of mean field decoupling, which renders the Hamiltonian quadratic and thus solvable (in a self-consistent manner). This procedure is most effectively carried out in $q$-space and we define the Fourier transform:
\begin{equation}\label{eq:hp-q-transform}
   a_{a,\mathbf{q}} = \frac{1}{\sqrt{N}}\sum_{\mathbf{q}} a_{ai} e^{-i\mathbf{q}\cdot \mathbf{r}_{i}},
\end{equation}
where $N$ denotes the number of unit cells in the system. The bosonic mean field decoupling then replaces terms like $a_{a,\mathbf{q}}a^\dagger_{b,\mathbf{q^\prime}}a^\dagger_{a,\mathbf{q^\prime}}a_{b,\mathbf{q-q^\prime-q^\dprime}}$ with terms of the form $ a_{a,\mathbf{q}}a^\dagger_{b,\mathbf{q^\prime}}\langle a^\dagger_{a,\mathbf{q^\prime}}a_{b,\mathbf{q-q^\prime-q^\dprime}} \rangle$, where $\langle\ldots\rangle$ denotes the thermal average and only terms diagonal in $\mathbf{q}$ yield a non-vanishing thermal average. We refer to appendix \ref{app:HP-comp} for details. Such a mean field treatment also imply that third order contributions can be ignored, since these can be transformed away similar to the first order terms. It is worth emphasizing though, that while the first order terms may always be ignored, the third order terms are only removed as a consequence of the mean field approximation. To proceed we consider a general system with $N_a$ magnetic sites in the primitive unit cell and we define the vector
\begin{equation}
    \mathbf{a}_{\mathbf{q}} =
    \begin{bmatrix}
        a_{a,\mathbf{q}} &  a^\dagger_{a,\mathbf{-q}} 
    \end{bmatrix}^T a\in [1,N_a], \label{eq:x_q}
\end{equation}
having length $2N_a$. The Hamiltonian in Eq. \eqref{eq:rot-heisenberg-model0} can then be written as
\begin{align}\label{eq:Hmodel_hp}
    \mathcal{H} &= \mathcal{E}_0 + \sum_{\mathbf{q}} \mathbf{a}^\dagger_{\mathbf{q}} \textrm{H}^\textrm{HP}_{\mathbf{q}} \mathbf{a}_{\mathbf{q}} \\
    \textrm{H}^\textrm{HP}_{\mathbf{q}} &=\textrm{H}_{\mathbf{q}}^0 + \textrm{H}_{\mathbf{q}}^1 \label{eq:H_hp}.
\end{align}
where $\mathcal{E}_0$, $\textrm{H}_{\mathbf{q}}^0$ and $\textrm{H}_{\mathbf{q}}^1$ are given by
\begin{align}
    \mathcal{E}_0 &= -\frac{1}{2}\sum_{abij} \mathbf{v}^T_a \textrm{J}^\prime_{abij}\mathbf{v}_b\Big( S_a S_b + \frac{S_a+S_b}{2}\Big) \\
    \textrm{H}_{\mathbf{q}}^0 &= -\frac{1}{2}
    \begin{bmatrix}
        \textrm{A}^0_{ab\mathbf{q}} - \textrm{C}^0_{ab} & \textrm{B}^0_{ab\mathbf{q}} \\
        \textrm{B}^{0 \dagger}_{ab\mathbf{q}}& \textrm{A}^{0 *}_{ab\mathbf{-q}} - \textrm{C}^0_{ab}
    \end{bmatrix}\label{eq:H0_q},\\
    \textrm{H}_{\mathbf{q}}^1 &= \frac{1}{2}
    \begin{bmatrix}
          \textrm{A}^1_{ab\mathbf{q}}- \textrm{C}^1_{ab\mathbf{q}} & \textrm{B}^1_{ab\mathbf{q}} \\
        \textrm{B}^{1 \dagger}_{ab\mathbf{q}} & \textrm{A}^{1 *}_{ab-\mathbf{q}} - \textrm{C}^{1 *}_{ab\mathbf{q}}
    \end{bmatrix} \label{eq:H1_q},
\end{align}
respectively and $\textrm{J}^\prime_{abij}$ is defined in Eq.\eqref{eq:Jp_abij} below. 
The individual entries in these matrices are defined in Eq. \eqref{eq:HP-comp1}-\eqref{eq:HP1-compL} in appendix \ref{app:HP-comp}. The first term $ \textrm{H}_{\mathbf{q}}^0$ provides the bare coupling between bosonic operators at second order and is equivalent to the Hamiltonian given in Ref. \cite{Toth2015}. The second term $ \textrm{H}_{\mathbf{q}}^1$ is temperature dependent and contains the thermal renormalization that originates from terms containing products of four bosonic operators in the original Hamiltonian.

While the A and C matrices determine the weight of operator products like $a^\dag_{a,\mathbf{q}}a_{b,\mathbf{q}}$ and $a_{a,\mathbf{q}}a_{b,\mathbf{q}}^\dag$ the B matrix is associated with products of the types $a^\dag_{a,\mathbf{q}}a^\dag_{b,\mathbf{q}}$ and $a_{a,\mathbf{q}}a_{b,\mathbf{q}}$, which do not conserve the boson number. % 
However, it is possible to perform a canonical transformation (para-diagonalization) that results in new operators $\alpha_{n,\mathbf{q}}$ that are linear combinations of the $a_{bj}$ and $a^\dag_{ai}$ such that non-conserving terms are eliminated while the bosonic commutator relations are conserved. 
This method was introduced by  Colpa \cite{COLPA1978327} and adapted for the HP bosonization of general single-$Q$ states in Ref. \onlinecite{Toth2015}. The new operators are thus written as
\begin{equation}\label{eq:T-def}
    \boldsymbol{\alpha}_\mathbf{q} = \textrm{T}_\mathbf{q} \mathbf{a}_\mathbf{q},
\end{equation}
with
\begin{equation}\label{eq:gen-BogVal}
    \boldsymbol{\alpha}_{\mathbf{q}} = 
    \begin{bmatrix}
       \alpha_{n,\mathbf{q}} & \alpha^{\dagger}_{n,\mathbf{-q}}
    \end{bmatrix}^T  n\in [1,N_a],
\end{equation}
and the Hamiltonian \eqref{eq:Hmodel_hp} can be diagonalized as 
\begin{equation}
    \textrm{E}_{\mathbf{q}} ={\textrm{T}_\mathbf{q}^{ \dagger-1}}\textrm{H}^\textrm{HP}_{\mathbf{q}}\textrm{T}_\mathbf{q}^{-1} \label{eq:ThT},
\end{equation}
with $\textrm{E}_{\mathbf{q}}$ being a diagonal matrix of the form
\begin{equation}
       \textrm{E}_{\mathbf{q}} =
   \frac{1}{2}\begin{bmatrix}
        \omega_{n,\mathbf{q}} & 0 \\
        0 & \omega_{n,\mathbf{-q}} 
   \end{bmatrix} .
\end{equation}
Applying this to Eq. \eqref{eq:Hmodel_hp} then gives
\begin{align}
    \mathcal{H} &= \mathcal{E}_0 + \frac{1}{2}\sum_{n\mathbf{q}}\Big(\omega_{n,\mathbf{q}} \alpha^\dagger_{n,\mathbf{q}}\alpha_{n,\mathbf{q}} \notag\\&\qquad\qquad\qquad+ \omega_{n,\mathbf{-q}} \alpha_{n,\mathbf{-q}}\alpha^\dagger_{n,\mathbf{-q}}\Big) \\
    &= \mathcal{E}_0 +  \sum_{n\mathbf{q}} \omega_{n,\mathbf{q}} ( \alpha^\dagger_{n,\mathbf{q}}\alpha_{n,\mathbf{q}}+\frac{1}{2}). \label{eq:H-hp-final}
\end{align}
The bosonic operators $\alpha_{n,\mathbf{q}}$ therefore define a new ground state that satisfies $\alpha_{n,\mathbf{q}}|0\rangle=0$ with an energy $E_0=\mathcal{E}_0 + \frac{1}{2}\sum_{n,\mathbf{q}}\omega_{n,\mathbf{q}}\le E_0^\mathrm{Cl}$ where $E_0^\mathrm{Cl}$ is the classical energy associated with the spiral state \eqref{eq:spin-spiral-gs}. The magnon energies $\omega_{n,\mathbf{q}}$ are renormalized at higher temperatures by the fourth order magnon interactions that are treated through a temperature-dependent mean field decoupling in $\textrm{H}^1_\mathbf{q}$ \eqref{eq:H1_q}. This implies that solutions at finite temperatures must be obtained in a self-consistent manner and the remaining task is to determine the transformation $\textrm{T}_\mathbf{q}$ in Eq. \eqref{eq:T-def}. It was shown by Colpa \cite{COLPA1978327} that this can be accomplished using the Cholesky decomposition
\begin{equation}
    \textrm{H}^\textrm{HP}_{\mathbf{q}} = \textrm{K}^\dagger_{\mathbf{q}} \textrm{K}_{\mathbf{q}},
\end{equation}
and the matrix
\begin{equation}
       \mathcal{I} =
   \begin{bmatrix}
        \textrm{I} & 0 \\ 0 & -\textrm{I}
   \end{bmatrix} 
\end{equation}
where $\textrm{I}$ is the identity matrix of size $N_a\times N_a$. If we then let $\textrm{U}_{\mathbf{q}}$ denote the unitary matrix that diagonalizes $\textrm{K}_{\mathbf{q}}\mathcal{I}\textrm{K}^\dagger_{\mathbf{q}}$:
\begin{equation}
    \mathcal{I}\textrm{E}_{\mathbf{q}} = \textrm{U}^{\dagger}_{\mathbf{q}}\big[\textrm{K}_{\mathbf{q}}\mathcal{I}\textrm{K}^\dagger_{\mathbf{q}}\big] \textrm{U}_{\mathbf{q}},
\end{equation}
the transformation matrix can be obtained as
\begin{equation}
    \textrm{T}_\mathbf{q} = \textrm{K}^{-1}_{\mathbf{q}} \textrm{U}_{\mathbf{q}}\textrm{E}^{1/2}_{\mathbf{q}}.
\end{equation}

\subsubsection{Magnetization}
Once we calculate the magnon energies, the sublattice magnetization at a given temperature can be obtained from the total magnon occupation numbers. In particular
\begin{equation}
    \langle S_a^z \rangle = S_a - \Phi_a .
\end{equation}
with
\begin{align}
    \Phi_a =&  \frac{1}{N}\sum_{\mathbf{q}} \langle a^\dagger_{a,\mathbf{q}} a_{a,\mathbf{q}} \rangle \notag \\
    =& \frac{1}{N}\sum_{\mathbf{q}} \langle \mathbf{a_q} \mathbf{a_q}^\dagger \rangle_{a + N_a, a + N_a} \\
    =& \frac{1}{N}\sum_{\mathbf{q}} \big[\textrm{T}_\mathbf{q} \langle \boldsymbol{\alpha}_{\mathbf{q}} \boldsymbol{\alpha}^\dagger_{\mathbf{q}} \rangle \textrm{T}^{\dagger}_\mathbf{q} \big]_{a + N_a, a + N_a}
\end{align}
and
\begin{equation}
   \langle \boldsymbol{\alpha}_{\mathbf{q}} \boldsymbol{\alpha}^\dagger_{\mathbf{q}} \rangle =
   \begin{bmatrix}
        1+n_\textrm{B}(\omega_{n,\mathbf{q}}) &  0 \\
        0 & n_\textrm{B}(\omega_{n,\mathbf{q}})
   \end{bmatrix}
\end{equation}
where $n_\textrm{B}(\omega_{n,\mathbf{q}})$ are Bose occupation factors.

For $T=0$, the classical spiral ground state \eqref{eq:spin-spiral-gs} has a site magnetization of $S_a$ by construction. However, except for the case of ideal ferromagnets, the ground state defined by $\alpha_{n,\mathbf{q}}|0\rangle=0$ will have $\langle0|S_a^{\dprime z}|0\rangle<S_a$. 
At elevated temperatures, the magnon dispersion is renormalized due to the magnon interactions, and the sub-lattice magnetization is reduced. 
As will be shown below, the theory breaks down before the sublattice magnetization vanishes because the system exhibits an instability and yields negative magnon energies. This constitutes a  pathological feature of the model, which may make estimates of the critical temperature somewhat dubious. Nevertheless, the derivative of the sub-lattice magnetization with respect to temperature approaches $-\infty$ at the point where the magnon energies become negative and it is natural to define that temperature as the critical point where magnetic order is lost. This will be the interpretation adopted in this work.

\subsection{Green's function method}\label{sec:Gfn}
The Green's function method was introduced by Bogoliubov and Tyablikov and is commonly referred to as the Random Phase Approximation (RPA). Superficially the physics and approximations here appear rather similar to the renormalized (HP) spinwave theory, but there are subtle and important differences as will be explained below. Originally developed for ferromagnetic order, the approach may be derived from the equation of motion for a single component of the dynamic transverse susceptibility (the Green's function $G^{+-}$) \cite{tyablikov1959, tahir1962}. For the general case of spiral order structures it turns out that one needs two components of the transverse susceptibility, which are coupled through the equations of motion. We define these in the local rotated frame such that
\begin{align}
    G^{+-}_{abij}(t) &= -i\theta(t) \langle [S^{\dprime +}_{ai}(t), S^{\dprime -}_{bj}]\rangle \equiv \langle\langle  S^{\dprime +}_{ai}(t) ;S^{\dprime -}_{bj} \rangle \rangle \label{eq:G_+-},\\
    G^{--}_{abij}(t)  &= -i\theta(t)\langle[ S^{\dprime -}_{ai}(t), S^{\dprime -}_{bj}] \rangle \equiv \langle\langle S^{\dprime -}_{ai}(t) ;S^{\dprime -}_{bj} \rangle \rangle, \label{eq:G_--}
\end{align}
where $\theta(t)$ is the Heaviside step-function. The temporal Fourier transforms are represented as
\begin{align}
    G_{abij}^{\pm-}(\omega) &= \int_{-\infty}^\infty dt\, e^{i(\omega+i\epsilon)t}\langle\langle S^{\dprime \pm}_{ai}(t) ; S^{\dprime -}_{bj} \rangle \rangle  \notag \\&\equiv  \langle\langle S^{\dprime \pm}_{ai} ; S^{\dprime -}_{bj} \rangle \rangle_\omega, \label{eq:G(w)}
\end{align}
where $\epsilon$ is an infinitesimal positive frequency. Since these Green's functions constitute different components of the dynamic spin susceptibility, the complex poles yields the energy eigenvalues of the quasi-particle excitations and can interpreted as magnon energies. It will be convenient to introduce the generalized commutator
\begin{equation}\label{eq:Gen-comm}
    [S^{\dprime\pm}_{ai}, \mathbf{S}^\prime_{bj}] = \mathbf{\Gamma}^\pm_{ai} \delta_{ab}\delta_{ij},
\end{equation}
as well the rotated exchange tensor:
\begin{equation}\label{eq:Jp_abij}
    \textrm{J}^\prime_{abij} = \textrm{R}^T_i \textrm{J}_{abij}\textrm{R}_j.
\end{equation}
Using this, the equations of motion for the Green's functions can be written as
\begin{align}
    (\omega + i\epsilon) G^{+-}_{abij}(\omega) &= 2\langle S^{\dprime z}_a \rangle \delta_{ab} \delta_{ij} \notag \\ & -\frac{1}{2}\sum_{ck} \big( \langle \langle \mathbf{S}^{\prime T}_{ck}  \textrm{J}^\prime_{caki} \mathbf{\Gamma}^+_{ai} ; S^{\dprime -}_{bj} \rangle \rangle_\omega \notag \\&+  \langle \langle \mathbf{\Gamma}^{+T}_{ai}  \textrm{J}^\prime_{acik} \mathbf{S}^\prime_{ck}  ; S^{\dprime -}_{bj} \rangle \rangle_\omega\big) \label{eq:eqofmotionGfn+},\\
    (\omega + i\epsilon) G^{--}_{abij}(\omega) &=  -\frac{1}{2}\sum_{ck}  \big( \langle \langle \mathbf{S}^\prime_{ck} \textrm{J}^\prime_{caki} \mathbf{\Gamma}^-_{ai} ; S^{\dprime -}_{bj} \rangle \rangle_\omega \notag\\& +\langle \langle \mathbf{\Gamma}^{-T}_{ai}  \textrm{J}^\prime_{acik} \mathbf{S}^\prime_{ck}  ; S^{\dprime -}_{bj} \rangle \rangle_\omega\big) \label{eq:eqofmotionGfn-}.
\end{align}
The right hand sides depend on higher order Green's functions (involving more than two spin operators) and render the equations difficult to solve. 
 To proceed, we employ the RPA which approximates the higher order Green's functions as products of spin expectation values and two-operator Green's functions. For ferromagnets magnetized along the $z$-direction, the basic idea is to approximate the individual fluctuations in $S^{z}_{ai}$ as the ensemble average ($\langle S^{z}_{a} \rangle$). For antiferromagnets and spiral ordered states it is sensible to decouple the fluctuations along the local magnetization direction, but since we are working in the locally rotated frame, this is simply represented by the $S_{ai}^{\dprime z}$ operators and we take:
\begin{equation}\label{eq:RPA-decoupling}
    \langle \langle S^{\dprime z}_{ck} S^{\dprime \pm}_{ai} ; S^{\dprime -}_{bj}  \rangle \rangle_\omega \approx \langle S^{\dprime z}_c \rangle \langle \langle S^{\dprime \pm}_{ai} ; S^{\dprime -}_{bj}  \rangle \rangle_\omega.
\end{equation}
If the rotated exchange tensor contains certain off-diagonal elements, other types of Greens function will enter the equations. However, as discussed in appendix \ref{app:Gfn} it is consistent to neglect Green's functions of the type $\langle\langle S^{\dprime\pm}_{ai}S^{\dprime \pm}_{ck}; S^{\dprime -}_{bj}\rangle\rangle_\omega$, $\langle\langle S^{\dprime\pm}_{ai}S^{\dprime \mp}_{ck}; S^{\dprime -}_{bj}\rangle\rangle_\omega$ and $\langle\langle S^{\dprime z}_{ai}S^{\dprime z}_{ck}; S^{\dprime -}_{bj}\rangle\rangle_\omega$ within the RPA. After applying Eq. \eqref{eq:RPA-decoupling} to Eqs. \eqref{eq:eqofmotionGfn+} and \eqref{eq:eqofmotionGfn-}, the equations may then be represented as a generalized eigenvalue problem:
\begin{align}\label{eq:eqofmotionGfn}
    \big((\omega + i\epsilon)\textrm{I} - \textrm{H}^\textrm{RPA}_\mathbf{q} \big) \textrm{G}_\mathbf{q}(\omega) &= \zeta,
\end{align}
where the RPA Hamiltonian, $\textrm{H}^\textrm{RPA}_\mathbf{q}$ is given by 
\begin{align}
    \textrm{H}^\textrm{RPA}_\mathbf{q} &=
    \begin{bmatrix}
        -\textrm{A}_{ab\mathbf{q}} +  \textrm{C}_{ab} & -\textrm{B}_{ab\mathbf{q}} \\
         \textrm{B}^\dagger_{ab\mathbf{q}} & \textrm{A}^*_{ab\mathbf{-q}} - \textrm{C}_{ab} 
    \end{bmatrix}\label{eq:H_rpa} \\
    \textrm{A}_{ab\mathbf{q}} &= \frac{\langle S^{\dprime z}_a \rangle}{2}\mathbf{u}^{T}_a \textrm{J}^{\prime}_{ab\mathbf{q}}\mathbf{u}^*_b\\
    \textrm{B}_{ab\mathbf{q}} &= \frac{\langle S^{\dprime z}_a \rangle}{2}  \mathbf{u}^{T}_a \textrm{J}^{\prime}_{ab\mathbf{q}}\mathbf{u}_b\\
    \textrm{C}_{ab} &= \delta_{ab}\sum_c \langle S^{\dprime z}_c \rangle  \mathbf{v}^T_a \textrm{J}^{\prime}_{ac\mathbf{0}}\mathbf{v}_c  \\
    \textrm{J}^\prime_{ab\mathbf{q}} &= \sum_i \textrm{J}^\prime_{ab0i}  e^{i\mathbf{q}\cdot \mathbf{r}_i}\label{eq:Jprime_q}
\end{align}
and
\begin{align}
    \textrm{G}_\mathbf{q}(\omega) &=
    \sum_i 
    \begin{bmatrix}
         G_{ab0i}^{+-}(\omega) & G_{ab0i}^{--}(\omega) 
    \end{bmatrix}^T e^{i\mathbf{q}\cdot\mathbf{r}_i}\\
    \zeta &=
    \begin{bmatrix}
         2\langle S^{\dprime z}_a \rangle \delta_{ab}  & 0  
    \end{bmatrix}^T.
\end{align}
We note that while $\textrm{H}^\textrm{RPA}_\mathbf{q}$ is a $2N_a\times2N_a$ matrix both $\textrm{G}_\mathbf{q}(\omega)$ and $\zeta$ are  $2N_a\times N_a$ matrices.
The Green's function \eqref{eq:eqofmotionGfn} can then finally be written as
\begin{align}\label{eq:G_q-result}
    \big[\textrm{G}_\mathbf{q}(\omega)\big]_{\alpha b} =\sum_{\eta \gamma} U_{\alpha \eta \mathbf{q}} \Big[&\frac{1}{\omega- \omega_{\eta,\mathbf{q}} + i\epsilon}\Big]  U^{-1}_{\eta  \gamma \mathbf{q}}[\zeta]_{\gamma b}, \\& \quad(\alpha,\eta) \in [1,2N_a]\notag
\end{align}
where $U_{\alpha \eta \mathbf{q}}$ is the matrix that diagonalizes the Hamiltonian \eqref{eq:H_rpa} and $\omega_{\eta,\mathbf{q}}$ are the corresponding eigenvalues. The eigenvalues are pairwise equal with opposite signs and the magnon energies are taken as the positive values. We note that the appearance of poles at both positive and negative eigenenergies are expected from the Lehmann representation of the Green's functions.
At finite temperatures the magnon energies become renormalized and the equations have to be solved self-consistently, since the Hamiltonian \eqref{eq:H_rpa} depends on thermal averages that can be expressed in terms of eigenstates and eigenenergies.

\subsubsection{Magnetization}
The magnetization for spin-1/2 systems can be obtained directly by solving the Green's functions defined in Eqs. \eqref{eq:G_+-}-\eqref{eq:G_--} \cite{beyondrpa} and using the fluctuation-dissipation theorem. For $S>1/2$ the relation between magnetization and Green's function is more complicated, but may be derived in terms of an auxiliary Green's function, which replaces the role of $G^{+-}$ \cite{Callen1963}. This results in the local magnetization being expressed as
\begin{equation}\label{eq:magnetization}
    \langle S_a^{\dprime z}  \rangle = \frac{(S_a-\Phi_a)(1+\Phi_a)^{2S_a+1} + (S_a+1+\Phi_a)\Phi_a^{2S_a+1}}{(1+\Phi_a)^{2S_a+1} - \Phi_a^{2S_a+1}}.
\end{equation}
where
\begin{equation}\label{eq:magnon-number-gfn}
    \Phi_a = \frac{1}{N}\sum_{\eta,\mathbf{q}} U_{a \eta \mathbf{q}} n_\textrm{B}(\omega_{\eta,\mathbf{q}}) U^{-1}_{\eta a \mathbf{q}}.
\end{equation}
In contrast to the case of HP bosons the Bose factor $n_\textrm{B}(\omega_{\eta,\mathbf{q}})$ appearing here should be regarded as a mathematical construct, since the sum runs over all $\omega_{\eta,\mathbf{q}}$, which contains both positive and negative eigenvalues of Eq. \eqref{eq:H_rpa}.
The magnetization has to be obtained self-consistently for a given temperature as the magnon energies are renormalized by the magnetization. In analogy with the HP method, the ground state magnetization for antiferromagnets and helimagnets are lower than $S_a$ as the magnon number defined in Eq. \eqref{eq:magnon-number-gfn} may be finite at $T=0$. At finite temperatures, the magnon dispersion softens and gradually becomes flat at the critical temperature. This implies a continuous decrease of magnetization as the critical temperature is approached from below and is in sharp contrast to the HP method that exhibits a discontinuity in the magnetization at the critical temperature.
\begin{figure}
    \centering
    \includegraphics[scale=0.33]{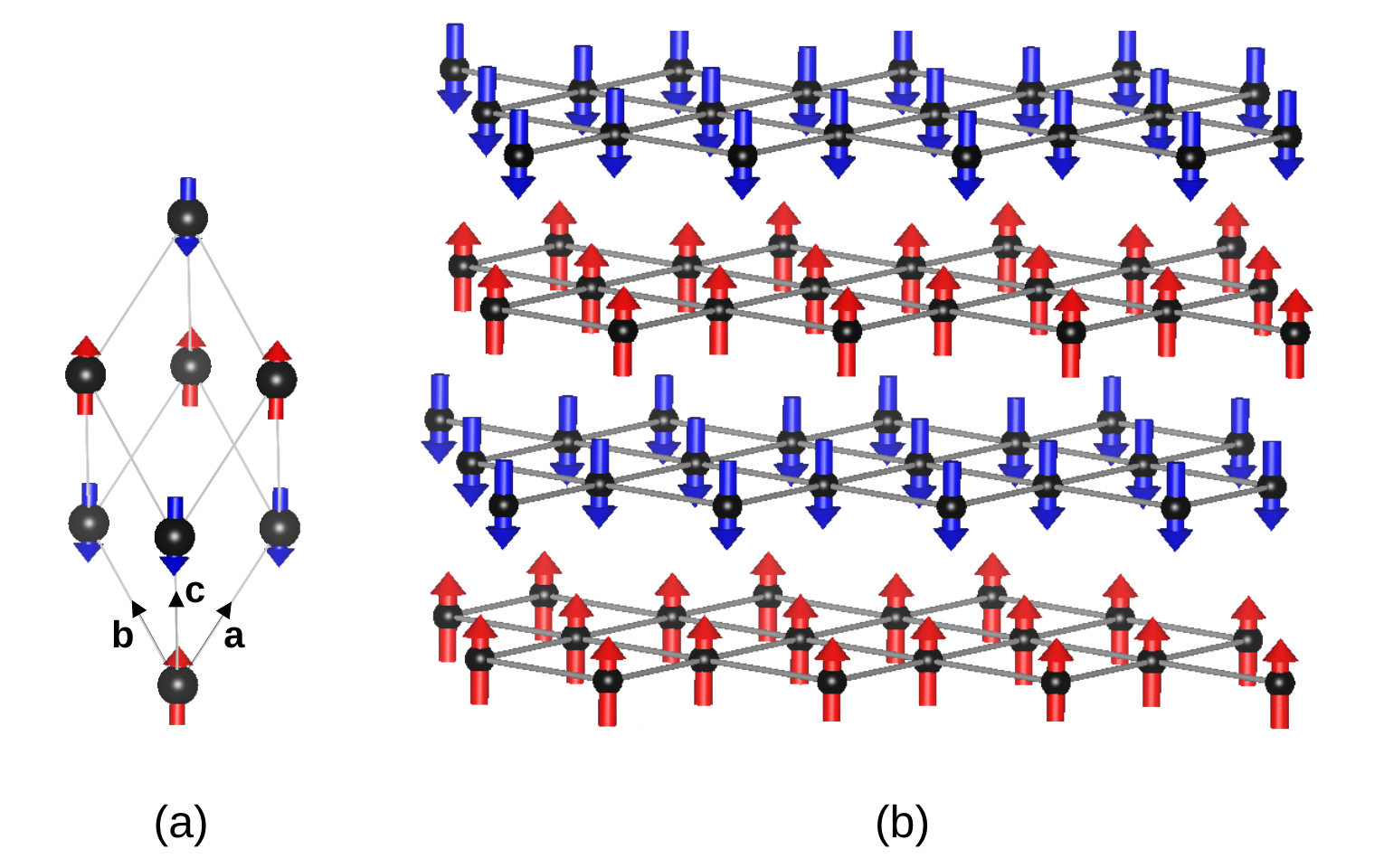}
    \caption{(a) The crystallographic fcc unit cell of NiO and MnO (only showing the magnetic atoms). (b) The magnetic order can be visualized as ferromagnetic hexagonal sheets with alternating moments along the [111] direction. This corresponds to  $\mathbf{Q}=[1/2,1/2,1/2]$ ordering of the unit cell in (a).}
    \label{fig:fcc-cartoon}
\end{figure}
\begin{figure*}[htbp]
    \includegraphics[scale=0.75]{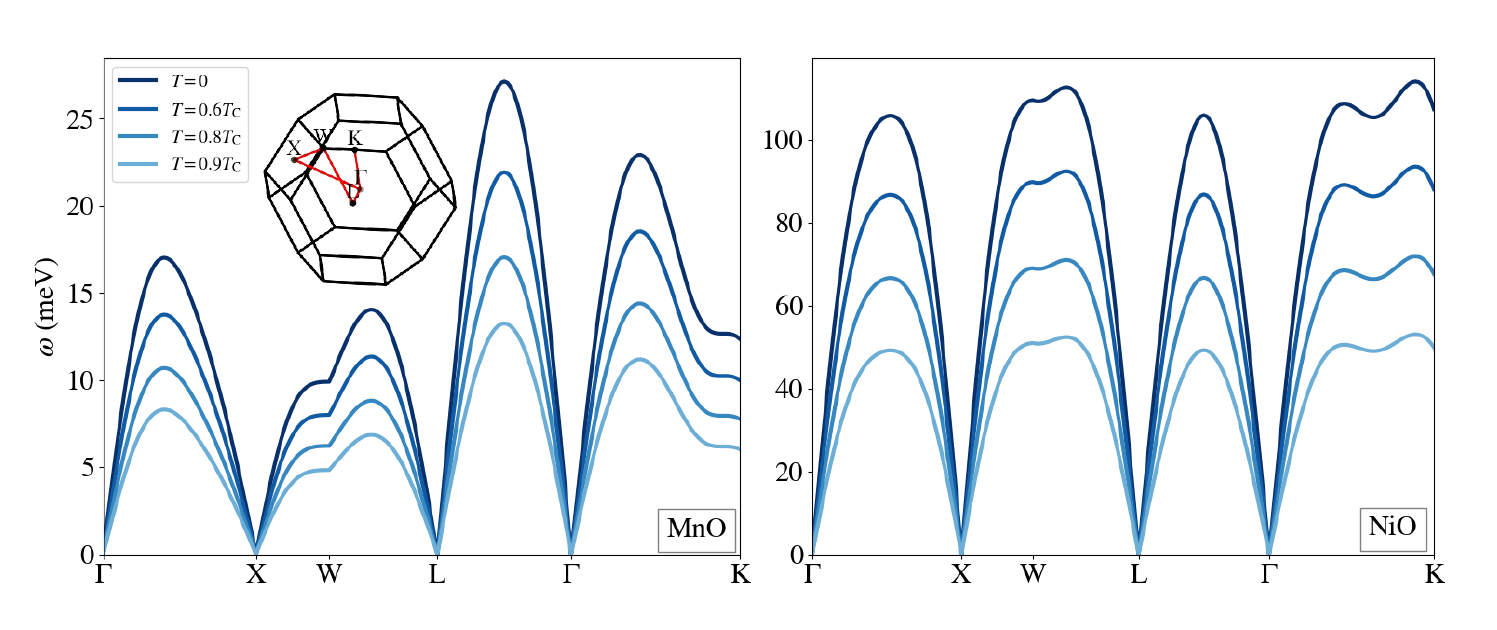}
    \caption{The  magnon dispersions of MnO and NiO calculated along high symmetry lines of the fcc unit cell shown in Fig. \ref{fig:fcc-cartoon}(a). The $T=0$ results were obtained with HP and the finite temperature results were obtained with RPA. $T_\mathrm{C}$ is the Néel temperature predicted by RPA.}
    \label{fig:fcc-disp}
\end{figure*}
For an isotropic model with a spiral ground state on a Bravais lattices, a closed analytical expression for the critical temperature can be obtained, which generalizes the well known ferromagnetic case \cite{beyondrpa}. Expanding Eqs. \eqref{eq:magnetization}-\eqref{eq:magnon-number-gfn} in the limit of $\langle S^{\dprime z}_a \rangle\rightarrow 0$, the critical temperature can be expressed as
\begin{align}\label{eq:T_RPA}
    k_\mathrm{B}T^\textrm{RPA}_\mathrm{C} &= \frac{S(S+1)}{3}\mathcal{F}^{-1}\\
    \mathcal{F} &= \frac{2}{N} \sum_\mathbf{q} \frac{|J_\mathbf{q} + \Tilde{J}_\mathbf{q} - 2\Tilde{J}_\mathbf{0}|}{|J_\mathbf{q} + \Tilde{J}_\mathbf{q} - 2\Tilde{J}_\mathbf{0}|^2 - |J_\mathbf{q} - \Tilde{J}_\mathbf{q}|^2}
\end{align}
with
\begin{align}
    J_\mathbf{q} =&  \sum_i J_{0i} e^{i\mathbf{q}\cdot \mathbf{r}_i}, \\
    \Tilde{J}_\mathbf{q} =& \sum_i J_{0i} \mathrm{cos}(\mathbf{Q}\cdot\mathbf{r}_i)  e^{i\mathbf{q}\cdot \mathbf{r}_i}
\end{align}
and $J_{0i}$ are the isotropic exchange coefficients.

\subsection{Mean Field Approximation}\label{sec:MFA}
Until now, we have discussed methods based on collective spin fluctuations. However, the simplest approach to include thermal effects is arguably Weiss mean field theory, where the interactions between magnetic moments are treated as an effective field. For ferromagnetic order, the effective field becomes uniform on the entire lattice, but in general one may define the effective field such that it is always aligned with the {\it local} magnetic moment. For a general spin spiral, 
the mean field Hamiltonian can be written as
\begin{align}
    \mathcal{H}_\textrm{MF} &= -\sum_{ai}\mathbf{S}_{ai}\cdot \mathbf{B}^\mathrm{eff}_{ai}\notag\\
    &=-\sum_{ai}S_{ai}^{\dprime z}B^{\dprime z}_{ai}
\end{align}
where the effective field is:
\begin{equation}
    B^{\dprime z}_{ai} =B^{\dprime z}_{a}= \sum_{bj} \langle S_{b}^{\dprime z} \rangle \mathbf{v}^T_a\textrm{J}_{ab0j}^{\prime} \mathbf{v}_b.
\end{equation}
The thermal average of the spin at site $a$ is then given by
\begin{equation}\label{eq:mfa-mag}
    \langle S_a^{\dprime z}\rangle = \frac{\sum_{m=-S_a}^{S_a} m e^{mB^{\dprime z}_{a}/k_\mathrm{B} T}}{\sum_{m=-S_a}^{S_a} e^{ mB^{\dprime z}_{a}/k_\mathrm{B}T}},
\end{equation}
and expanding this expression in small average magnetization then yields
\begin{align}\label{eq:T_MFA}
    k_\mathrm{B}T^\textrm{MFA}_\mathrm{C} \langle S^{\dprime z}_a \rangle &= \frac{S_a(S_a+1)}{3} \sum_{bj} \langle S^{\dprime z}_b \rangle \mathbf{v}^T_a\textrm{J}_{ab0j}^{\prime} \mathbf{v}_b\notag\\
    &=\frac{S_a(S_a+1)}{3} \sum_{b} \langle S^{\dprime z}_b \rangle \mathbf{v}^T_a\textrm{J}_{ab\mathbf{0}}^{\prime} \mathbf{v}_b,
\end{align}
where $\textrm{J}_{ab\mathbf{0}}^{\prime}$ was defined in Eq. \eqref{eq:Jprime_q}. For the case of equivalent spins the critical temperature is therefore given by $\frac{S_a(S_a+1)}{3k_\mathrm{B}} \sum_{b} \mathbf{v}^T_a\textrm{J}_{ab\mathbf{0}}^{\prime} \mathbf{v}_b$, which is independent of $a$. For non-equivalent spins Eq. \eqref{eq:T_MFA} defines a consistency relation in the limit of infinitesimal magnetization. However, the condition may be satisfied at finite magnetization as well, which would always happen below the critical temperature and we thus take the critical temperature to be the largest eigenvalue of the matrix 
$\frac{S_a(S_a+1)}{3k_\mathrm{B}} \sum_{b} \mathbf{v}^T_a\textrm{J}_{ab\mathbf{0}}^{\prime} \mathbf{v}_b$. It is finally noted that for an isotropic system with spins defined by Eq. \eqref{eq:spin-spiral-gs} and exchange interactions $J_{abij}$, we have
\begin{align}
 \mathbf{v}^T_a\textrm{J}_{ab\mathbf{0}}^{\prime} \mathbf{v}_b &= \sum_i J_{ab0i} \mathbf{v}^T_a\textrm{R}_\mathbf{\hat{x}}(\mathbf{Q}\cdot\mathbf{r}_i)\mathbf{v}_b \notag\\
 &=\sum_i J_{ab0i}\cos(\mathbf{Q}\cdot\mathbf{r}_i+\phi_b-\phi_a)\notag\\
 &=\mathcal{R}\textit{e}[J_{ab\mathbf{Q}}e^{i(\phi^b-\phi^a)}],
\end{align}
where we took $\mathbf{v}_a =  \begin{bmatrix}0,&-\mathrm{sin}\phi_a,& \mathrm{cos}\phi_a\end{bmatrix}$ consistent with a spiral in the $yz$-plane. This shows that the critical temperature is largely (precisely for a Bravais lattice) determined by $J_{ab\mathbf{Q}}$ (see also Appendix \ref{app:Classical ground state}).

\section{Results}\label{sec:results}
Above we introduced a generalized framework for calculating the Néel temperatures of spin spirals using three different methods. Here we benchmark the methods by calculating Néel temperatures of various specific materials and compare the results to experimental values. We consider MnO, NiO, MnF$_2$, Cr$_2$O$_3$, Fe$_2$O$_3$ and Ba$_3$NbFe$_3$Si$_2$O$_{14}$ (BNFS), which represent different examples of ordering vectors as well as differences in intra-cell order. The exchange constants of all compounds have been extracted from inelastic neutron scattering experiments and are known to accurately reproduce the measured magnon dispersions at low temperatures. We have tabulated these in Tab. \ref{table:J_n}. In addition to the RPA, HP and MFA, which all represent quantum mechanical treatments, we also compare with classical Monte Carlo simulations.
%, which is a fully correlated method, but neglect quantum effects.
\begin{figure}
    \centering
    \includegraphics[scale=0.35]{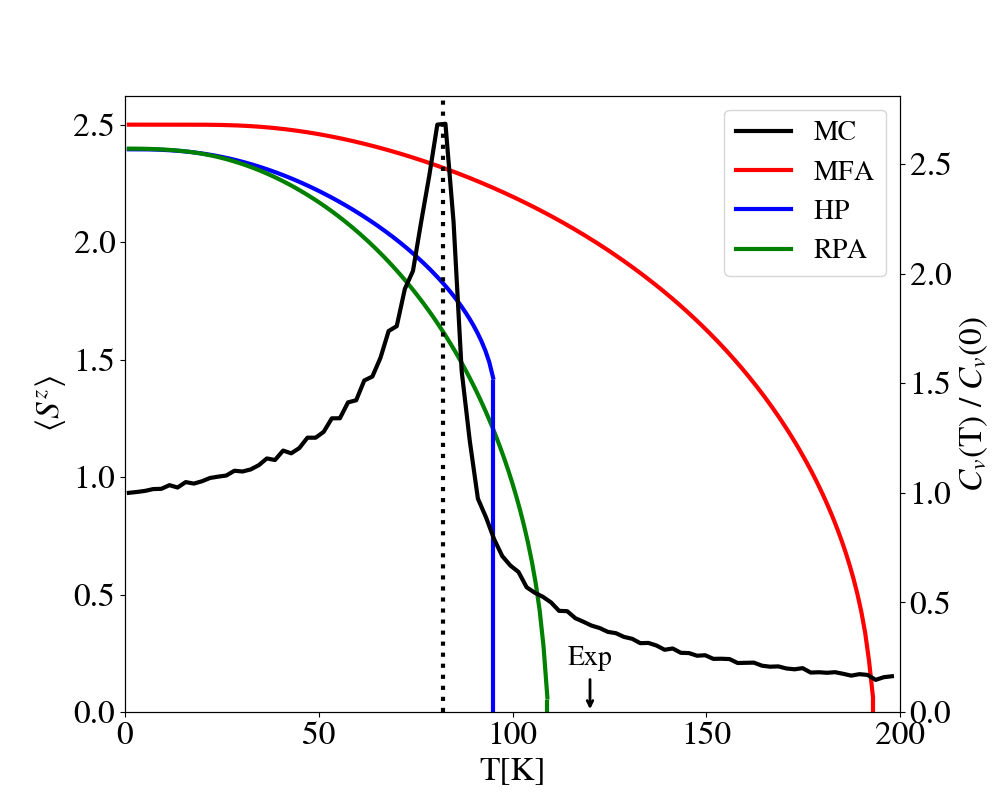}
    \caption{Magnetization and heat capacity of MnO calculated as a function of temperature from MC (heat capacity), MFA, HP and RPA. The experimental Néel temperature is indicated on the \textit{x}-axis.}
    \label{fig:mno-sz}
\end{figure}

\subsection{MnO and NiO}
\begin{figure}
    \centering
    \includegraphics[scale=0.35]{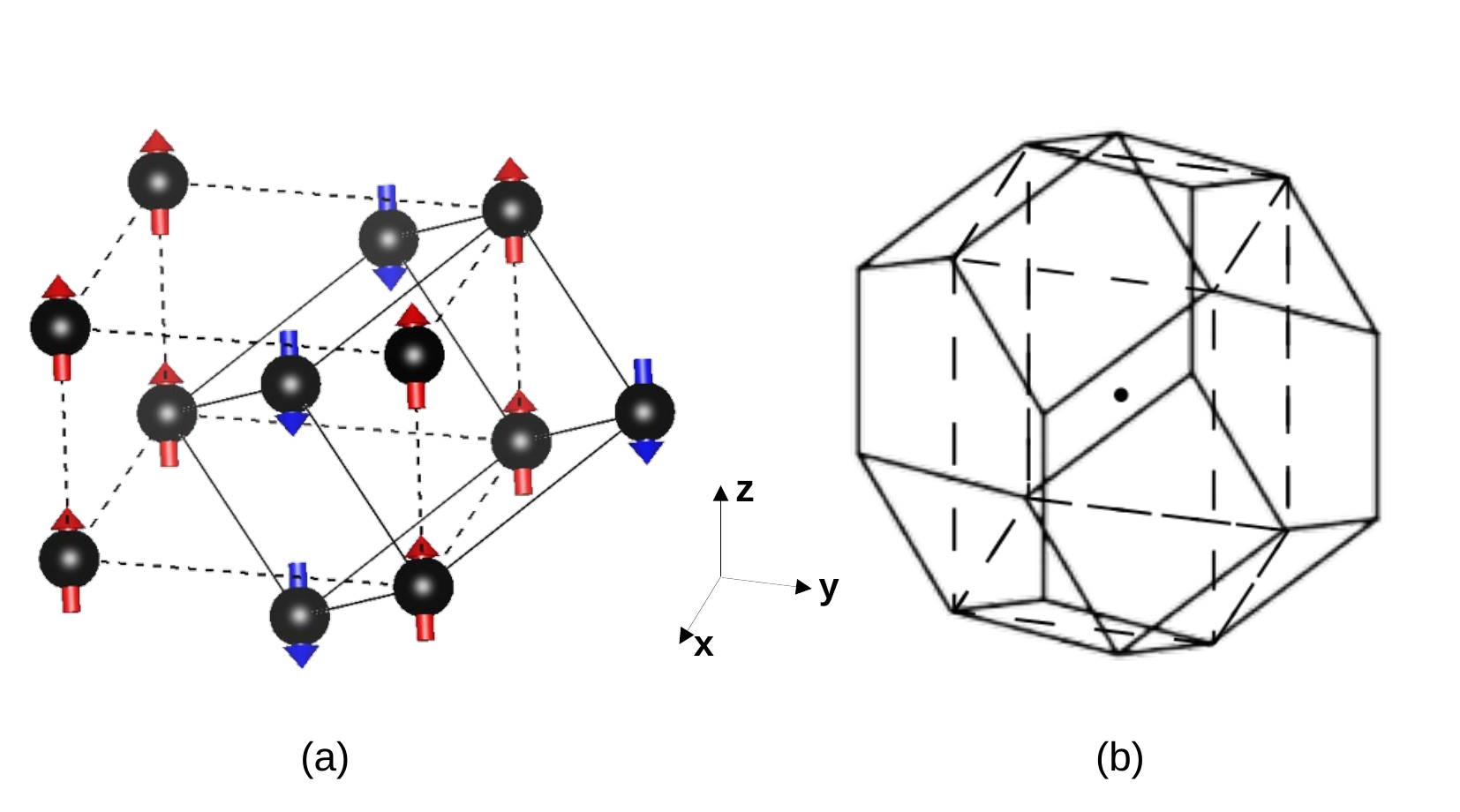}
    \caption{(a) The primitive unit cell (solid lines) and the altermagnetic unit cell (dashed lines) of the tetragonal body centered magnetic lattice of MnF$_2$. (b) The first Brillouin zone of the primitive (solid) and the altermagnetic (dashed) unit cell of MnF$_2$.}
    \label{fig:MnF2-cartoon}
\end{figure}
\begin{figure*}[htbp]
    \centering
    \includegraphics[scale=0.74]{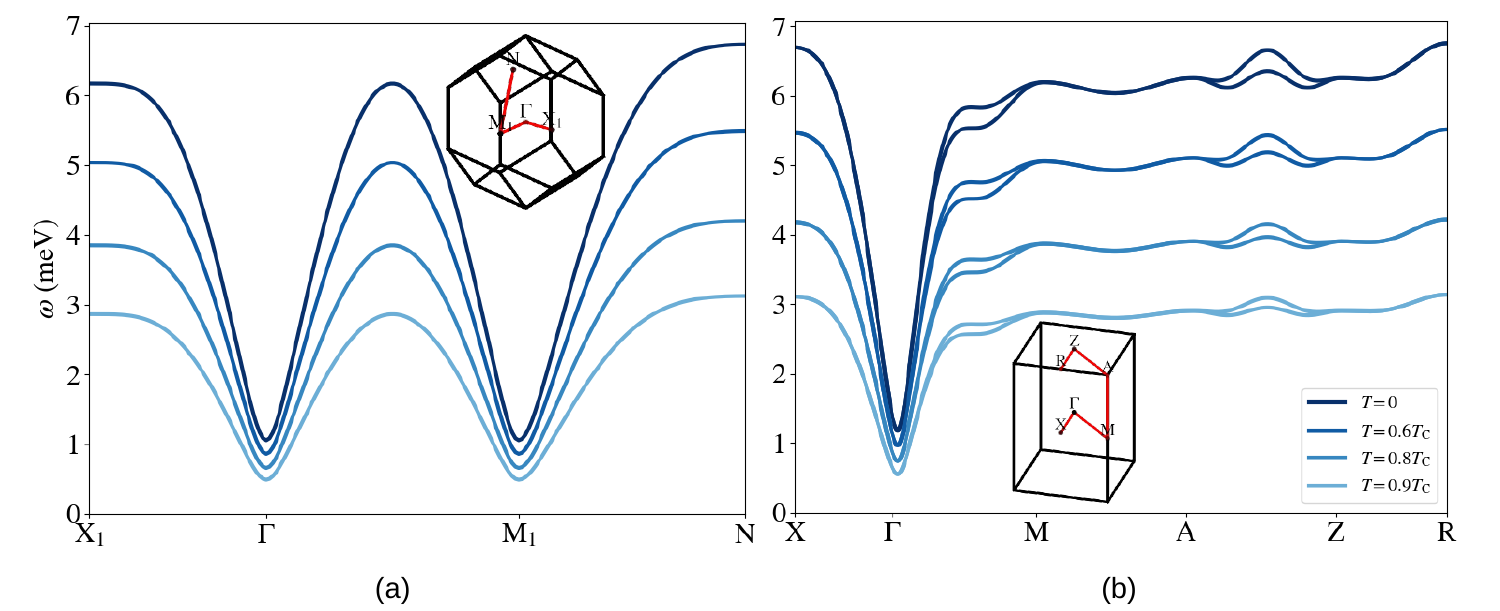}
    \caption{(a) The magnon dispersion of MnF$_2$ calculated along high symmetry lines in the primitive cell of the tetragonal body centered lattice. The $T=0$ results were obtained with HP and the finite temperature results were obtained with RPA. $T_\mathrm{C}$ is the Néel temperature predicted by RPA. (b) Same as (a) except the  calculations were performed in the two-atom tetragonal unit cell, where we have allowed for an altermagnetic magnon band splitting.}
    \label{fig:MnF2-disp}
\end{figure*}
At low temperatures MnO and NiO form rock salt lattices with a slight trigonal distortion \cite{Roth1958-dist}. Here we neglect the trigonal distortion and the magnetic atoms of the rock salt structures, thus compose fcc lattices. The magnetic order in both compounds is collinear antiferromagnetic with ferromagnetic planes alternating along the (111) direction. The ground state spin is $S=5/2$ and $S=1$ for MnO and NiO respectively. The magnetic order is illustrated in Fig. \ref{fig:fcc-cartoon} along with the crystallographic unit cell of the fcc lattice. In order to represent the magnetic order one may use a trigonal unit cell containing two magnetic atoms along the (111) direction (with antialigned spins) or we may simply use the crystallographic unit cell and impose $\mathbf{Q}=[1/2,1/2,1/2]$ order through the boundary conditions. MnO and NiO have been widely studied in the past as prototypical examples of antiferromagnets \cite{PhysRevB.66.064434,PhysRevB.91.125133,Olsen2017} and the experimental exchange constants (taken from Ref. \cite{Lines1965MnO} and Ref.  \cite{Hutchings1972NiO} respectively) are listed in Table \ref{table:J_n}.

The magnon dispersion relations of MnO and NiO calculated in the crystallographic unit cell are shown in Fig. \ref{fig:fcc-disp}. In addition to the zero temperature dispersions calculated using HP, we also show the thermally renormalized magnon energies using RPA at $T=0.6T_\mathrm{C}$, $T=0.8T_\mathrm{C}$ and $T=0.9T_\mathrm{C}$.
%In addition to the zero temperature dispersions (which are the same for HP and RPA) we also show the thermally renormalized magnon energies using RPA at $T=0.6T_\mathrm{C}$, $T=0.8T_\mathrm{C}$ and $T=0.9T_\mathrm{C}$. 
As these are calculated in the unit cell with a single magnetic site, the dispersion contains a single magnon branch. In contrast, the magnetic unit cell (containing two magnetic atoms) yields two degenerate magnon branches with gapless states at the trigonal high symmetry points $\Gamma$ and F. The gapless points at $\Gamma$, X and L in Fig. \ref{fig:fcc-disp} are the ones that downfold to $\Gamma$ and F in the magnetic unit cell. %

The softening of the magnon dispersion at elevated temperatures shown in Fig. \ref{fig:fcc-disp} is a direct consequence of the thermal reduction of magnetic moments. In Fig. \ref{fig:mno-sz} we show the magnetic moment of a Mn site in MnO as a function of temperature calculated from RPA, MFA and HP. At $T=0$, the correlated ground state gives rise to a magnetization that satisfies $\langle S^z\rangle < S$, which are seen to be captured by both RPA and HP. The MFA, in contrast, is referenced directly to the classical ground state and has  $\langle S^z\rangle = S$ by construction at $T=0$. The critical temperature can be estimated as the point where the magnetization vanishes. 
As mentioned above, RPA and MFA continuously decrease to zero, while the magnetization in HP exhibits a discontinuity at the point where the derivative (as a function of temperature) approaches $-\infty$. This happens because the thermally renormalized dispersion in HP becomes negative at finite magnetization. In addition, we also show the heat capacity calculated from classical Monte Carlo simulations, which exhibits a sharp peak at the Néel temperature. Compared with the experimental Néel temperature (120 K), we see that RPA provides the best estimate (106 K), whereas HP and classical MC underestimate it (95 K and 82 K respectively) and the MFA severely overestimates it (192 K). We note that the underestimated value obtained from a classical treatment is expected - even for a high spin compound such as MnO. From either the MFA Eq. \eqref{eq:T_MFA} or RPA Eq. \eqref{eq:T_RPA} expressions for the critical temperature $T_\mathrm{C}$, it is evident that the classical analogue is obtained by the replacement $(S+1)\rightarrow S$ and is related to the quantum results by $T_\mathrm{C}^\mathrm{QM}=\frac{S+1}{S}T_\mathrm{C}^\mathrm{Cl}$. We can thus obtain a ``poor man's quantum corrected MC" result by multiplying the classical Néel temperature by $(S+1)/S$. Such an approach was also applied in Ref. \cite{Alaei2023} to improve the predictions of classical MC. For Mn ions in the $S=5/2$ state this gives a factor of 1.4 and a ``corrected" MC Néel temperature of 115 K, which is in good agreement with the experimental value of 120 K. The calculated and experimental Néel temperatures are summarized in Tab. \ref{table:Tc}. 

\begin{figure*}[htbp]
    \centering 
    \includegraphics[scale=0.6]{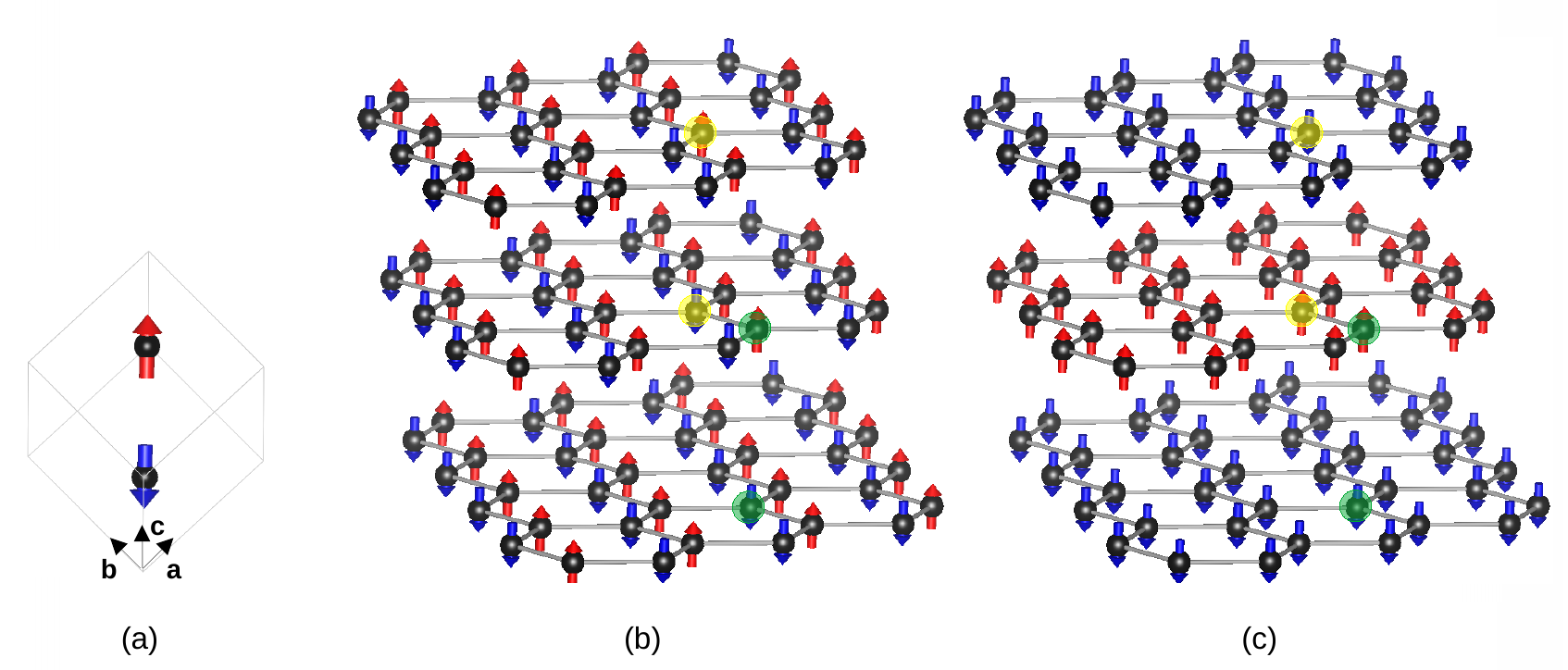}
    \caption{(a) The minimal  unit cell of Cr$_2$O$_3$ and Fe$_2$O$_3$ (only showing magnetic atoms) with anti-aligned magnetic moments at the two sites. The magnetic order can be visualized as (buckled) honeycomb sheets with ABC 
    stacking along [111] direction for (b) Cr$_2$O$_3$(Néel ordered sheets) and (c) Fe$_2$O$_3$(ferromagnetic sheets). Atoms that are directly on top of each other are marked with colors green and yellow. The  buckling of the honeycomb layers has been removed for visualization.}
    \label{fig:cor-cartoon}
\end{figure*}
\subsection{MnF$_2$}
MnF$_2$ crystallizes in an tetragonal body-centered structure \cite{doi:10.1021/ja01650a005} with two Mn atoms (with antialigned 5/2 spins) in the tetragonal unit cell and lattice parameters given by \textit{a}=\textit{b}=4.8 {\AA} and \textit{c}=3.31 {\AA}. It comprises a canonical example of a uniaxial collinear antiferromagnet \cite{RevModPhys.25.338} and the exchange constants and single-ion anisotropy constant used here have been extracted from Ref. \cite{Okazaki1964}. These are listed in Tab. \ref{table:J_n}. Note that we are including a single-ion anisotropy parameter ($A$) for this compound, which is given by half the $\textrm{J}_{aaii}^{zz}$ component of the exchange tensor. 
Neglecting the ligands, one may represent the magnetic sites as a Bravais lattice with $\mathbf{Q}=[1/2,1/2,1/2]$ and apply a procedure similar to the cases of NiO and MnO and calculate the single-site magnon dispersion within this unit cell (subject to the $Q$-ordering). The result of this is shown in Fig. \ref{fig:MnF2-disp}(a). However, in contrast to the cases of NiO and MnO, the F atoms in the structure implies that the crystallographic unit cell cannot be chosen as the primitive tetragonal bcc cell. Moreover, since the spin compensation is provided by a four-fold rotation symmetry rather than a combination of inversion and time-reversal the compound can be classified as altermagnetic \cite{MnF2-Altermagnet} . This implies that it is unjustified to cast the system into a single site unit cell and that the two magnon branches in the required two-site unit cell(shown in Fig.\ref{fig:MnF2-cartoon}(a), may be non-degenerate on certain paths in the Brillouin zone. The splitting of magnon branches is, however, expected to be small and has not yet been verified experimentally. Moreover it is not clear if such splitting (and the required altermagnetic representation in a two-site unit cell) has any consequences for the thermal properties. In order to include the altermagnetic splitting, we have calculated the magnon spectrum from first principles using the magnetic force theorem \cite{Durhuus_2023}. This calculation reveals that the maximal magnon splitting is on the order of 5 {\%} of the magnon band width. Such splitting can be included in the model here if the experimental exchange constants are augmented by a fifth nearest interaction that is allowed to have different values in different directions (but the different values being constrained by four-fold symmetry). In Fig. \ref{fig:MnF2-disp}(b), we show this modified magnon dispersion evaluated in a two-site unit cell and observe the expected magnon splitting on the $\Gamma$-M and A-Z paths. We have evaluated the Néel temperature for both the antiferromagnetic ($\mathbf{Q}=[1/2,1/2,1/2]$) state and the altermagnetic ($\mathbf{Q}=[0,0,0]$) state and get RPA values of 71 K in both cases. In Tab. \ref{table:Tc} we list the Néel temperatures obtained from the different methods and again note that the RPA is in excellent agreement with the experimental value of 67 K. Again the HP and MC simulations severely underestimate the experimental value (52 K and 42 K respectively), but the MC result can be significantly improved by performing the phenomenological quantum correction (multiplying by $(S+1)/S$), which yields 59 K.

\subsection{Cr$_2$O$_3$ and Fe$_2$O$_3$}
We now proceed with the slightly more complex structures of Cr$_2$O$_3$ and Fe$_2$O$_3$, which crystallize in the corundum structure with two formula units in the minimal unit cell.
The magnetic atoms can be regarded as a stack of buckled honeycomb lattices with each layer being Néel ordered in the case of Cr$_2$O$_3$ and ferromagnetic ordered in the case of Fe$_2$O$_3$ as illustrated in Fig. \ref{fig:cor-cartoon}. Ignoring the non-magnetic atoms, both lattice geometries can be represented by a rhombohedral unit cell with two magnetic atoms with antialigned magnetic moments \cite{Samuelsen1969swcor} (see Fig. \ref{fig:cor-cartoon}) 
The difference between the two structures is then given by $\mathbf{Q}=[0,0,0]$ and $\mathbf{Q}=[1/2,1/2,1/2]$ for Cr$_2$O$_3$ and Fe$_2$O$_3$ respectively. We emphasize that in contrast to the altermagnetic case, this representation provides an exact account of the site-based magnetic model because all exchange constants connecting magnetic sites in the magnetic unit cell can be represented in the minimal two-site cell due to symmetry. 

\begin{figure*}[htbp]
    \includegraphics[scale=0.75]{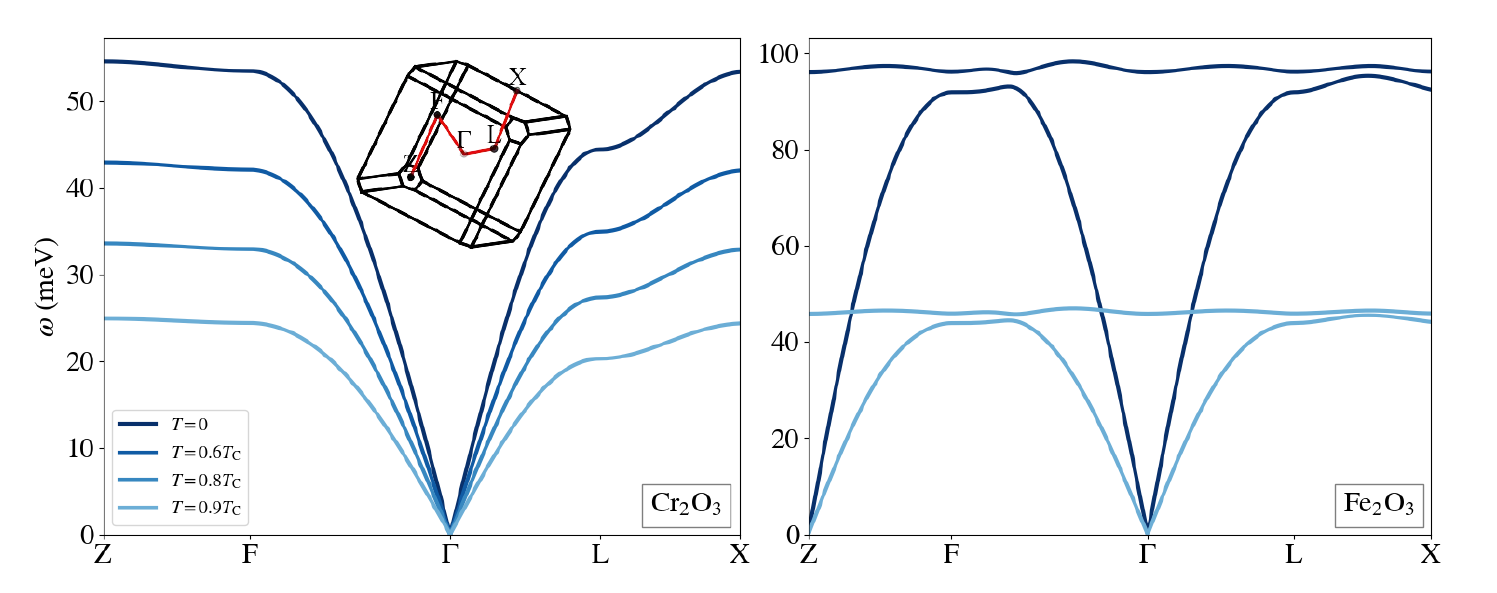}
    \caption{Magnon dispersions of Cr$_2$O$_3$ and Fe$_2$O$_3$ calculated along high symmetry lines of the primitive cell shown in  Fig.\ref{fig:cor-cartoon}(a). The $T=0$ results were obtained with HP and the finite temperature results were obtained with RPA. $T_\mathrm{C}$ is the Néel temperature predicted by RPA. For clarity, the results for Fe$_2$O$_3$ are only shown for $T=0$ and $T$=0.9$T_\mathrm{C}$.}%The dispersion is calculated using RPA at %$T=0$,$T=0.6T_\mathrm{C}$, $T=0.8T_\mathrm{C}$ and $T=0.9T_\mathrm{C}$ for Cr$_2$O$_3$ and at $T$=0 and $T$=0.9$T_\mathrm{C}$ for Fe$_2$O$_3$.}
    \label{fig:cor-disp}
\end{figure*}
The experimental exchange constants for Cr$_2$O$_3$ and Fe$_2$O$_3$ were obtained from Ref. \onlinecite{Samuelsen1969Cr2O3} and Ref. \onlinecite{Samuelson1970Fe2O3} respectively and are listed (converted to our conventions) in Tab. \ref{table:J_n}. The magnon dispersions for Cr$_2$O$_3$ and Fe$_2$O$_3$ calculated in the minimal unit cell exhibits two magnon branches corresponding to the two atoms in the unit cell and are displayed in Fig. \ref{fig:cor-disp}. Both structures lead to two magnon modes, but whereas the $\mathbf{Q}=[0,0,0]$ modes of Cr$_2$O$_3$ are exactly degenerate, the finite-$Q$ structure of Fe$_2$O$_3$ shows one flat and one dispersive mode. The Néel temperature of Fe$_2$O$_3$ is 960 K, which is more than a factor of three larger than Cr$_2$O$_3$ (see Tab. \ref{table:Tc}) and is a consequence of the larger magnon bandwidth and the larger magnetic moment. Again, RPA provides the best agreement with experiments for both compounds, whereas HP and MC underestimate the Néel temperatures, and MFA severely overestimates them.

\subsection{Ba$_3$NbFe$_3$Si$_2$O$_{14}$}
\begin{figure}[htbp]
    \centering
    \includegraphics[scale=0.28]{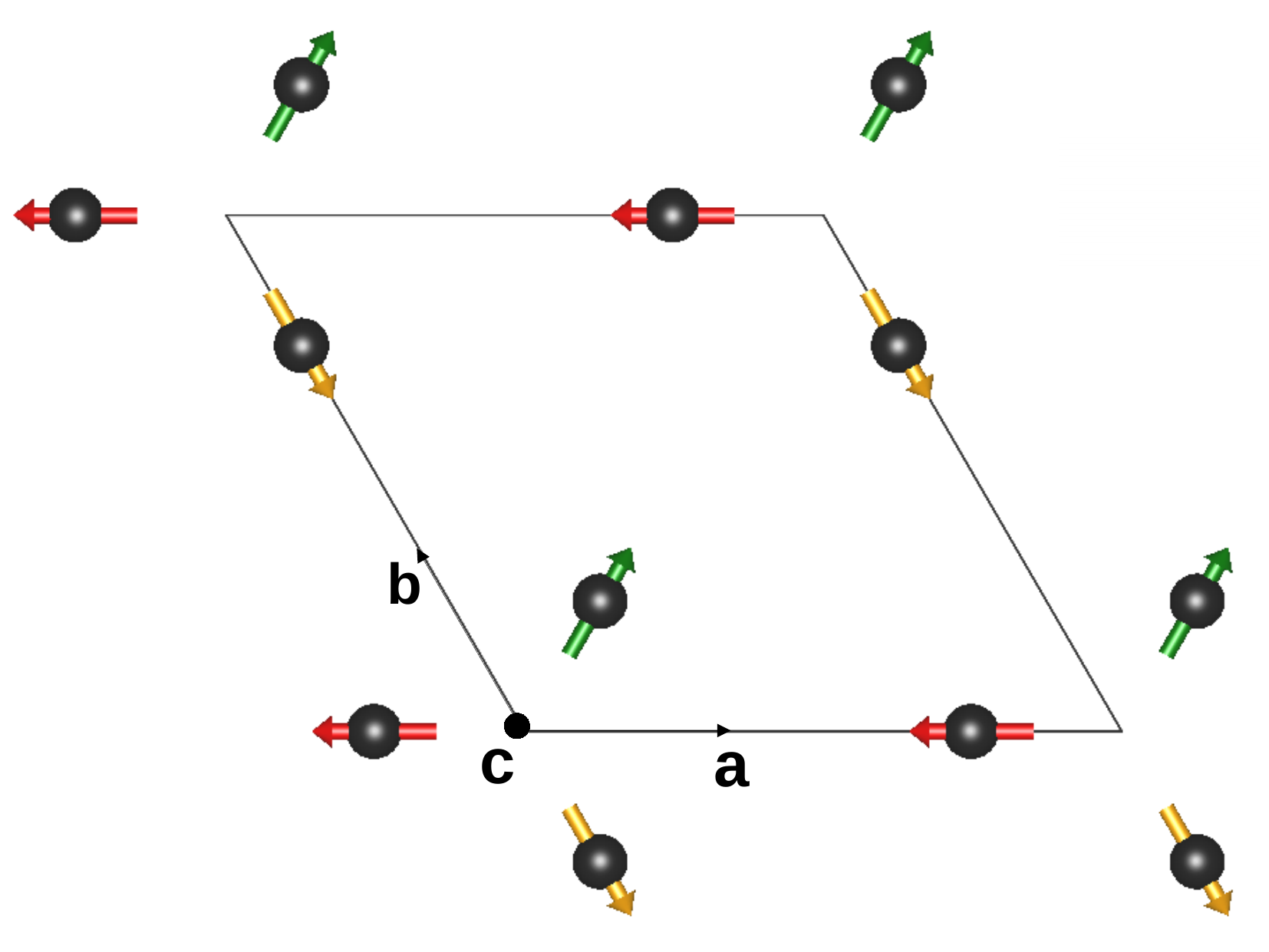}
    \caption{The magnetic structure of Ba$_3$NbFe$_3$Si$_2$O$_{14}$ in the \textit{ab}-plane. The arrangement of spins in the out-of-plane direction is (approximately) described by a $\mathbf{Q}=[0,0,1/7]$ ordering vector.}
    \label{fig:bnfs-cartoon}
\end{figure}

All the materials we considered until now have had collinear magnetic ground states and although there are scattered results in the literature on how to perform RPA on collinear antiferromagnets, these have been limited to specific lattice geometries \cite{Lee1967,Samuelsen1969swcor}. To exemplify the generality of the present method, we consider  
Ba$_3$NbFe$_3$Si$_2$O$_{14}$ (BNFS), which comprises an incommensurate single-$Q$ structure with three non-collinear magnetic moments in the minimal unit cell. 
The magnetic lattice is thus formed by Fe$_3$-trimers exhibiting a 120$^0$ arrangement of magnetic moments and the trimer units form hexagonal AA stacked planes. With this intracell arrangement the magnetic order is described by $\mathbf{Q}=[0,0,1/7]$ (ordering vector orthogonal to the hexagonal planes). The structure is illustrated in Fig. \ref{fig:bnfs-cartoon}. 
The corresponding exchange constants are taken from Ref. \onlinecite{Loire2011} and are listed in Tab. \ref{table:J_n}. As the DM interaction obtained from the fitting was very small (1\% of $|J_1|$), we neglect it here. We note that this system was also chosen in Ref. \cite{Toth2015} as an example of how to calculate the spinwave dispersion in non-collinear systems at $T=0$.

\begin{figure}
    \centering
        \includegraphics[scale=0.32]{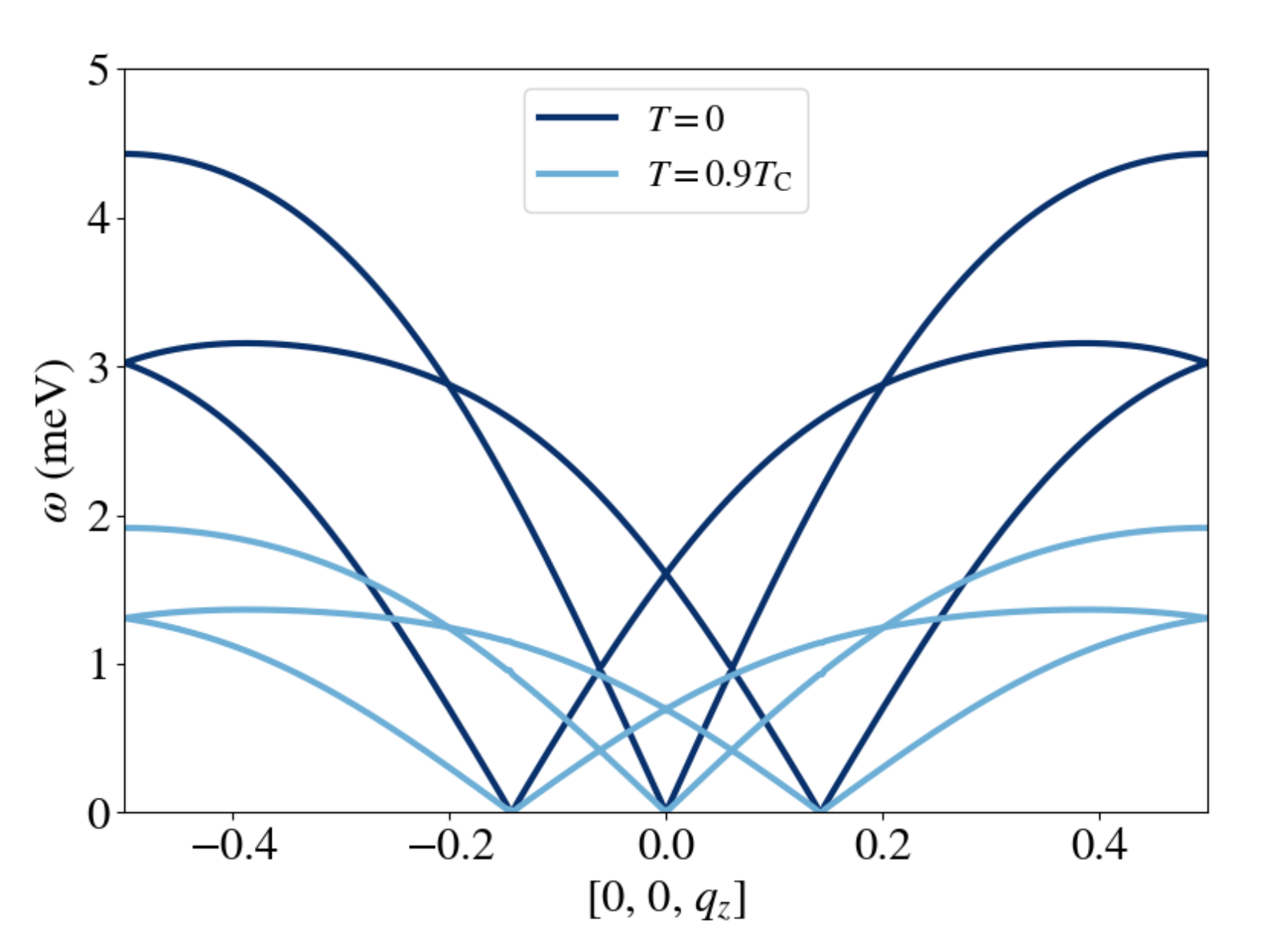}
        \caption{The magnon dispersion along the \textit{z}-direction of Ba$_3$NbFe$_3$Si$_2$O$_{14}$ at $T=0$(HP) and $T=0.9T_\mathrm{C}$(RPA).}%calculated from RPA.}
    \label{fig:bnfs-disp}
\end{figure}

The magnon dispersion of BNFS calculated along the \textit{z}-direction is shown in Fig. \ref{fig:bnfs-disp}. The three branches are a consequence of having three Fe atoms in the crystallographic unit cell and are in good agreement with the experimental dispersion in Ref. \onlinecite{Loire2011} (as required since the exchange constant was fitted to the data). The RPA and HP yield slightly different results at $T=0$ and here we show the HP results, which is identical to that reported in Ref. \cite{Toth2015}. We also show the $T=0.9T_\mathrm{C}$ dispersion, which exhibits strong thermal softening. In Tab. \ref{table:Tc} we list the Néel temperatures obtained with the different methods and we observe that the HP and MC methods again severely underestimates the Néel temperature. Similarly to the other compounds considered in this work, MFA overestimates the critical temperature and RPA offers the best agreement with the experimental Néel temperature.

\section{Conclusion and Outlook}
\begin{table}[t]
    \centering
    \begin{tabular}{ |p{1.5cm}||p{1.1cm}|p{1.1cm}|p{0.9cm}|p{0.9cm}|p{0.9cm}||p{0.9cm}|}
        \hline
        & $J_1$ & $J_2$ & $J_3$ & $J_4$ & $J_5$ & $A$\\
        \hline
        MnO \cite{Lines1965MnO} & -0.8617  &  -0.9479  & 0  &  0  & 0 & 0 \\
        NiO \cite{Hutchings1972NiO} & 1.37  &  -19.01  & 0  &  0  & 0 & 0\\
        MnF$_2$ \cite{Okazaki1964} & 0.056  &  -0.304  & 0  &  0  & 0 & 0.0182\\
        Cr$_2$O$_3$ \cite{Samuelsen1969Cr2O3} & -15.04 &  -6.82  & -0.155  &  0.034  & -0.379 & 0\\
        Fe$_2$O$_3$ \cite{Samuelson1970Fe2O3} &  1.034 &  0.2756  & -5.12  &  -3.99  & -0.172 & 0\\
        BNFS \cite{Loire2011} & -0.85  &  -0.24  & -0.053  &  -0.017  & -0.24 & 0\\
        \hline
    \end{tabular}
    \caption{The exchange constants used in this work. $J_n$ represents the $n^{th}$ nearest neighbor interaction for each of the materials. All values are in meV and were taking from experimental references (obtained by fitting to inelastic neutron scattering data). Note that some of the values are rescaled according to our conventions (see Eq. \eqref{eq:heisenberg-model}).}
    \label{table:J_n}
\end{table}
\begin{table}[t]
    \centering
    \begin{tabular}{ |p{1.2cm}||p{1.4cm}|p{1.2cm}|p{1.2cm}|p{1.2cm}|p{1.2cm}|}
        \hline
        & MC & MFA & RPA & HP & Expt \\
        \hline
        MnO & 82(114)  &  192  & 106  &  95  & 120 \\
        NiO & 310(620)  &  882  & 588  &  424  & 525 \\
        MnF$_2$ & 42(59)  &  88  & 71  &  52  & 67 \\
        Cr$_2$O$_3$ &   192(320)    &  530  & 292  &  200  & 307 \\
        Fe$_2$O$_3$ &  730(1022) &  1318  & 974  &  750  & 960 \\
        BNFS & 15(21)  &  41  & 30  &  16  & 27 \\
        \hline
    \end{tabular}
    \caption{Calculated critical temperatures (in \textrm{K}) of the materials considered in this work. For the classical MC results we also show (in brackets) the ``poor man's quantum corrected results" obtained by multiplying the classical value by $(S+1)/S$.}% In all cases, MC and HP are underestimates, MFA is an overestimate while RPA gives excellent agreement with experiments.}
    \label{table:Tc}
\end{table}
We have introduced two methods to calculate the thermal renormalization of dispersion relations and the Néel temperature of general single-\textit{Q} spin spirals. The first scheme was based on the renormalized spin wave theory using the HP bosonization scheme. This approach is well known for simple collinear systems, but a completely general framework for single-$Q$ spirals had not been developed prior to this work. Here we generalized the rotating frame of reference developed in Ref. \onlinecite{Toth2015} to include thermal effects through a mean field treatment of magnon-magnon interactions, which may used to calculate the Néel temperature. The second approach is referred to as the RPA and is based on the equation of motion for the dynamical Green's functions. Again, there are scattered results in the literature where such an approach has been applied to various simple antiferromagnets, but a general framework for single-$Q$ spirals was missing. Here we have applied a rotating frame approach to derive the RPA equations for general single-$Q$ spiral ground states, which may be used to calculate Néel temperatures within the RPA approximation.

%and the Green's function RPA. Using these methods, we calculated the magnon dispersions and critical temperature of 
We applied the methods to the cases of MnO, NiO, Cr$_2$O$_3$, Fe$_2$O$_3$ and Ba$_3$NbFe$_3$Si$_2$O$_{14}$ where the exchange constants were taken from fitted inelastic neutron scattering data. These materials represent various kinds of single-\textit{Q} spirals including single-site collinear order, two-site collinear order, altermagnetic order and three-site spiral order. 
%In the zero temperature limit, the HP method (which is equivalent to linear spin wave theory \cite{Toth2015}) predicts a dispersion relation identical to that of RPA. But at elevated temperatures, the renormalized magnon energies calculated using HP and RPA starts to diverge. 
The experimentally obtained magnon energies yield slightly different exchange constants when mapped to the HP dispersion compared to the RPA dispersion. This is because the RPA dispersion is scaled by $\langle S^{\dprime z}_a\rangle$ whereas HP is scaled by $S_a$. Thus, the exchange constants obtained by fitting to RPA will differ by a factor of $S_a/\langle S^{\dprime z}_a\rangle$ compared to HP. To avoid confusion, we have retained the exchange constants obtained by fitting experimental dispersion relations to HP in all calculations. However, we emphasize that for non-ferromagnetic systems, RPA implies a rescaling of exchange constants compared to a fit based on HP. Such rescaling would give exchange constants of slightly larger magnitude and is likely to yield more consistent results.
As we approach the phase transition, the HP magnon energies abruptly jump to unstable negative modes indicating a pathological behavior of the model, whereas the RPA magnon dispersion smoothly flattens to zero. The critical temperature of all the materials using these methods along with classical MC and MFA was benchmarked against the experimental Néel temperatures. Similar to the case of ferromagnetic order \cite{beyondrpa}, the Néel temperature using MFA is severely overestimated and classical MC yields much too low values - even for high spin systems such as MnO and MnF$_2$ (both having $S=5/2$) that are often considered ``classical". 
The most accurate predictions are provided by the RPA, which also comprises a rather simple framework compared to for example HP (with thermal renormalization) or classical MC (which requires careful convergence of iterations, system size etc). This is true for collinear antiferromagnets as well as non-collinear structures and RPA thus constitutes a simple method of choice for prediction of the Néel temperature given a set of exchange constants.

Except for the case of BNFS, we have only considered simple paradigmatic antiferromagnets where the magnon dispersion and exchange constants are well known. This choice was necessary in order to properly benchmark the methods against experimental Néel temperatures. Moreover, except for the case of MnF$_2$ we have only considered isotropic exchange. For MnF$_2$ we added a single-ion anisotropy interaction to obtain a magnon dispersion relation in agreement with the experimental one, but the anisotropy does not have much influence on the Néel temperature. It would be interesting to test the methodology on two-dimensional materials, where magnetic anisotropy is absolutely crucial for having strict long range order \cite{Olsen_2024}. 
Moreover, the effect of Dzyaloshinskii-Moriya interactions and Kitaev interactions has not been included in any of the materials studied here. These contributions may give rise to topological gaps in the magnon spectrum and it will be interesting to investigate how thermal effects renormalize such gaps. Finally, the methodology developed here easily lends itself to high throughput screening of new materials with exchange constants obtained from first principles and such screening studies could lead to the discovery of new high temperature magnetic materials.
\appendix
\section{Classical ground state in non-Bravais lattices} \label{app:Classical ground state}
The classical ground state order is the starting point for any calculation of excited states and in the following we will outline how it may be obtained from a set of exchange constants. We limit ourselves to isotropic exchange interactions and Eq. \eqref{eq:heisenberg-model} can then be reduced to
\begin{equation}\label{eq:heisenberg-model-iso}
    \mathcal{H} = -\frac{1}{2}\sum_{abij} J_{abij} \mathbf{S}_{ai}\cdot \mathbf{S}_{bj},
\end{equation}
where $\mathbf{S}_{ai}$ is the {\it classical} spin vector for site $a$ in unit cell $i$. 
The problem is thus to find the spin configuration that minimizes the classical energy given by Eq. \eqref{eq:heisenberg-model-iso} under the $N_a$ constraints
\begin{equation}\label{eq:s-constraint}
    \mathbf{S}_{ai} \cdot \mathbf{S}_{ai} = S^2_a.
\end{equation}
For $N$ unit cells, we define the Fourier transforms
\begin{align}
    \mathbf{S}_{ai} &= \frac{1}{\sqrt{N}}\sum_\mathbf{q} \mathbf{S}_{a\mathbf{q}} e^{i\mathbf{q}\cdot\mathbf{r}_i},\label{eq:S_fourier}\\
    J_{ab\mathbf{q}} &=\sum_i J_{ab0i}e^{i\mathbf{q}\cdot\mathbf{r}_i}.
\end{align}
For the case of a Bravais lattice (one atom in the unit cell and $N_a=1$) is it well known that the solution is a planar spin spiral \onlinecite{yosida2010theory} characterized by an ordering vector $\mathbf{Q}$, 
and there have been several attempts to generalize the method to non-Bravais lattices \cite{Lyons1960, LITVIN1974205,Schmidt2022}. 

In our opinion the simplest way to proceed is to formulate the problem in terms of Lagrange multipliers and we replace the ``strong" constraint \eqref{eq:s-constraint} by the weaker one:
\begin{align}
    \sum_{ai} \mathbf{S}_{ai} \cdot \mathbf{S}_{ai} &= \sum_{a\mathbf{q}} \mathbf{S}_{a\mathbf{q}}\cdot \mathbf{S}_{a\mathbf{q}}=\sum_a NS_a^2 \label{eq:w-constraint}.
\end{align}
We thus have to minimize the Lagrangian
\begin{align}
   \mathcal{L} =& -\frac{1}{2}\sum_{ab\mathbf{q}} J_{ab\mathbf{q}} \mathbf{S}_{a\mathbf{q}} \cdot \mathbf{S}_{b\mathbf{-q}}  \notag\\
   &+\frac{\lambda}{2} \sum_{a} \Big(\sum_\mathbf{q}\mathbf{S}_{a\mathbf{q}} \cdot \mathbf{S}_{a\mathbf{-q}} - NS_a^2\Big), 
\end{align}
where $\lambda$ is a Lagrange multiplier to be determined. The Lagrangian has an extremum when 
\begin{align}\label{eq:JqSqlSq}
    \sum_b J_{ab\mathbf{q}} \mathbf{S}_{b\mathbf{q}} = \lambda \mathbf{S}_{a\mathbf{q}}.
\end{align}
That is, $\lambda$ must be an eigenvalue of $J_{ab\mathbf{q}}$. There are $N$ such solutions for $\lambda$ corresponding to the different $\mathbf{q}$-vectors. However, for a given solution the corresponding extremal state can only have a single $\mathbf{q}$-vector with non-vanishing $\mathbf{S}_{a\mathbf{q}}$. Applying this to Eq. \eqref{eq:heisenberg-model-iso} with the weak constraint \eqref{eq:w-constraint} gives 
\begin{align}
    \mathcal{H} = -\frac{\lambda N}{2}\sum_a S_a^2.
\end{align}
Thus while any single-$Q$ state constitutes an extremal point (with different possibilities for the $\mathbf{S}_{a\mathbf{q}}$ being determined by the eigenvectors of $J_{ab\mathbf{q}}$), the minimum energy is obtained by finding the $\mathbf{q}$-vector that maximizes the largest eigenvalue of $J_{ab\mathbf{q}}$:
\begin{equation}
    \lambda = \textrm{max}_\mathbf{q}(\mathcal{J}^{n_\textrm{max}}_\mathbf{q}).
\end{equation}
A rather general approach for obtaining the eigenstates has been discussed in Ref.  \cite{Schmidt2022}, but we choose to focus on the situation  where there exists a single-$Q$ state such that the largest eigenvalues of the exchange tensor are identical at $\pm\mathbf{Q}$:
\begin{equation}\label{eq:J_nQ}
    \lambda = \mathcal{J}^{n_\textrm{max}}_\mathbf{\pm Q}.
\end{equation}
We then write the eigenvectors in terms of a basis-vector $\mathbf{s}$ such that
\begin{equation}
    \mathbf{S}_{a\mathbf{q}} =
    \begin{cases}
    0 & \text{$\mathbf{q}\neq\pm\mathbf{Q}$}\\
    v_a\mathbf{s}&\text{$\mathbf{q}=\mathbf{Q}$}\\
    v^*_a\mathbf{s}^*&\text{$\mathbf{q}=\mathbf{-Q}$}
    \end{cases},
\end{equation}
where $v_a$ is the corresponding eigenvector of $\mathcal{J}^{n_\textrm{max}}_\mathbf{Q}$, and $\mathbf{s}$ is a 3-dimensional complex vector. Applying the strong constraints \eqref{eq:s-constraint} and noting that only the $\pm\mathbf{Q}$ terms contribute to the sum \eqref{eq:S_fourier}, it is straighforward to show that the basis vector must satisfy
\begin{align}
\mathbf{s}\cdot\mathbf{s}&=0,\\
2|v_a|^2\mathbf{s}\cdot\mathbf{s}^*&=NS_a^2.
\end{align}
These constraints imply that the real and imaginary parts of $\mathbf{s}$ must be orthogonal and have the same norm. There are many (degenerate) solutions that satisfy this and here we choose one that leads to a spiral state in the $yz$-plane:
\begin{equation}
    \mathbf{s}=
    \begin{bmatrix}
        0& -i & 1
    \end{bmatrix}^T.
\end{equation}
This implies that
\begin{align}
|v_a|^2=\frac{NS_a^2}{4}
\end{align}
and writing $v_a=|v_a|e^{i\phi_a}$
the ground state spin configuration becomes
\begin{equation}
    \mathbf{S}_{ai} = S_a 
    \begin{bmatrix}
        0\\ \textrm{sin}(\mathbf{Q}\cdot \mathbf{r}_i + \phi_a)\\ \textrm{cos}(\mathbf{Q}\cdot \mathbf{r}_i + \phi_a) 
    \end{bmatrix}.
\end{equation}
In conclusion, for single-$Q$ spirals, the ordering vector is determined from the maximum eigenvalue of the exchange matrix \eqref{eq:J_nQ} and the intra-site angles between sites in the unit cell are determined by the phases of components of the associated eigenvector $v_a=|v_a|e^{i\phi_a}$.

\section{Holstein-Primakoff}\label{app:HP-comp}
Inserting Eq. \eqref{eq:Sprime-uv-hp} into Eq. \eqref{eq:rot-heisenberg-model0} yields
\begin{align}
\mathcal{H}\approx \mathcal{E}_0 + \mathcal{H}^1+\mathcal{H}^2+ \mathcal{H}^3+\mathcal{H}^4
\end{align}
where
\begin{align}
    \mathcal{E}_0 = -\frac{1}{2}\sum_{abij} \mathbf{v}^T_a \textrm{J}^\prime_{abij}\mathbf{v}_b( S_a S_b + \frac{S_a+S_b}{2}), 
\end{align}
and
\begin{align}\label{eq.H_1}
    \mathcal{H}^1 &=  -\frac{1}{2}\sum_{abij}\Big( S_b\sqrt{\frac{S_a}{2}}\big(\mathbf{u}^{*T}_a \textrm{J}^\prime_{abij}\mathbf{v}_b a_{ai} + \mathbf{u}^{T}_a \textrm{J}^\prime_{abij}\mathbf{v}_b a^\dagger_{ai}\big) \notag\\&\qquad\qquad+ S_a\sqrt{\frac{S_b}{2}}\big(\mathbf{v}^T_a \textrm{J}^\prime_{abij}\mathbf{u}^*_b a_{bj} + \mathbf{v}^T_a \textrm{J}^\prime_{abij}\mathbf{u}_b a^\dagger_{bj}\big)\Big)
\end{align}
is the part of the Hamiltonian linear in bosonic operators,
\begin{align}\label{eq:H_2}
    \mathcal{H}^2 &=  -\frac{1}{2} \sum_{abij} \Big(-\mathbf{v}^T_a \textrm{J}^\prime_{abij} \mathbf{v}_b\big(S_b \frac{a^\dagger_{ai} a_{ai}+ a_{ai} a^\dagger_{ai}}{2} \notag\\&\qquad\qquad\qquad\qquad\qquad\quad+ 
    S_a \frac{a^\dagger_{bj} a_{bj}+ a_{bj} a^\dagger_{bj}}{2}\big) \notag\\& +\frac{\sqrt{S_aS_b}}{2} \big( a_{ai} \mathbf{u}^{*T}_a \textrm{J}^\prime_{abij} \mathbf{u}^*_b a_{bj} + 
    a_{ai} \mathbf{u}^{*T}_a \textrm{J}^\prime_{abij} \mathbf{u}_b a^\dagger_{bj} \notag\\ &\qquad\qquad+
    a^\dagger_{ai} \mathbf{u}^T_a \textrm{J}^\prime_{abij} \mathbf{u}^*_b a_{bj} +
    a^\dagger_{ai} \mathbf{u}^T_a \textrm{J}^\prime_{abij} \mathbf{u}_b a^\dagger_{bj} \big)\Big),
\end{align}
is the part of the Hamiltonian quadratic in bosonic operators,
\begin{align}\label{eq:H_3}
    \mathcal{H}^3 &=  \frac{1}{2} \sum_{abij} \Big( S_b\sqrt{\frac{S_a}{2}}\big(\mathbf{u}^{*T}_a \textrm{J}^\prime_{abij}\mathbf{v}_b a^\dagger_{ai}a_{ai}a_{ai} \notag\\&\qquad\qquad\qquad\qquad+ \mathbf{u}^{T}_a \textrm{J}^\prime_{abij}\mathbf{v}_b a^\dagger_{ai}a^\dagger_{ai}a_{ai}\big) \notag\\&\qquad\qquad+ S_a\sqrt{\frac{S_b}{2}}\big(\mathbf{v}^T_a \textrm{J}^\prime_{abij}\mathbf{u}^*_b a^\dagger_{bj}a_{bj}a_{bj} \notag\\&\qquad\qquad\qquad\qquad+ \mathbf{v}^T_a \textrm{J}^\prime_{abij}\mathbf{u}_b a^\dagger_{bj}a^\dagger_{bj}a_{bj}\big)\Big),
\end{align}
is the part of the Hamiltonian cubic in bosonic operators and
\begin{align}\label{eq:H_4}
    &\mathcal{H}^4 = \frac{1}{2}\sum_{abij} \Bigl(-\mathbf{v}_a^T \textrm{J}^\prime_{abij} \mathbf{v}_b a^\dagger_{ai}a_{ai}a^\dagger_{bj}a_{bj} \notag\\&+\frac{\sqrt{S_a S_b}}{8}\Bigl( \mathbf{u}^{*T}_a \textrm{J}^\prime_{abij} \mathbf{u}^*_b \bigl( \frac{a_{ai}a^\dagger_{bj}a_{bj}a_{bj}}{S_b} + \frac{a^\dagger_{ai}a_{ai}a_{ai}a_{bj}}{S_a} \bigr) \notag \\&\qquad\qquad +\mathbf{u}^{*T}_a \textrm{J}^\prime_{abij} \mathbf{u}_b \big( \frac{a_{ai} a_{bj} a^\dagger_{bj} a^\dagger_{bj}}{S_b} + \frac{a_{ai}a_{ai} a^\dagger_{ai} a^\dagger_{bj}}{S_a} \big) \notag \\&\qquad\qquad+
    \mathbf{u}_a^T \textrm{J}^\prime_{abij} \mathbf{u}^*_b \bigl( \frac{a^\dagger_{ai}a^\dagger_{bj}a_{bj}a_{bj}}{S_b} + \frac{a^\dagger_{ai}a^\dagger_{ai}a_{ai}a_{bj}}{S_a} \bigr) \notag \\& \qquad\qquad+ \mathbf{u}_a^T \textrm{J}^\prime_{abij} \mathbf{u}_b \bigl( \frac{a^\dagger_{ai}a^\dagger_{bj}a^\dagger_{bj}a_{bj}}{S_b} + \frac{a^\dagger_{ai}a^\dagger_{ai}a_{ai}a^\dagger_{bj}}{S_a} \bigr) \Bigr) \Bigr),
\end{align}
is the part of the Hamiltonian quartic in bosonic operators. The linear part can be absorbed into the quadratic part (and a constant term) by a linear transformation of the bosonic operators and hence can be ignored for the sake of calculating excitation energies. From Eq. \eqref{eq:H_2} one may then directly read off Eq. \eqref{eq:H0_q} with
\begin{align}
    \textrm{A}^0_{ab\mathbf{q}} &= \frac{\sqrt{S_aS_b}}{2}  \mathbf{u}_a^T\textrm{J}^{\prime}_{ab\mathbf{q}} \mathbf{u}_b^* \label{eq:HP-comp1}\\
    \textrm{B}^0_{ab\mathbf{q}} &=\frac{\sqrt{S_aS_b}}{2}  \mathbf{u}_a^T\textrm{J}^{\prime}_{ab\mathbf{q}} \mathbf{u}_b \\
    \textrm{C}^0_{ab} &= \delta_{ab} \sum_c S_c \mathbf{v}^T_a \textrm{J}^\prime_{ac\mathbf{0}} \mathbf{v}_c,
\label{eq:HP0-compL}
\end{align}
and $\textrm{J}^{\prime}_{ab\mathbf{q}}$ given by Eq. \eqref{eq:Jprime_q}. This result is equivalent to the one found by Toth and Lake \cite{Toth2015}. The lowest order magnon-magnon interactions, however, are contained in $\mathcal{H}^3$ and $\mathcal{H}^4$, which can only be treated approximately. The idea is to take the quartic terms in bosonic operators and approximate them by quadratic terms weighted by expectation values. In order to exemplify the approach we consider the quartic term $a^\dagger_{a,\mathbf{q}} a^\dagger_{a,\mathbf{q^\prime}} a_{a,\mathbf{q^\dprime}} a^\dagger_{b,\mathbf{-q-q^\prime+q^\dprime}}$ (obtained by Fourier transforming the last term of Eq. \eqref{eq:H_4}),
for which the bosonic mean field decoupling yields
\begin{align}
    \frac{1}{N}\sum_{\mathbf{q,q^\prime,q^\dprime}}&a^\dagger_{a,\mathbf{q}} a^\dagger_{a,\mathbf{q^\prime}} a_{a,\mathbf{q^\dprime}}  a^\dagger_{b,\mathbf{-q-q^\prime+q^\dprime}}e^{i(\mathbf{q+q^\prime-q^\dprime})\cdot\mathbf{r}_j} \notag\\
    \approx \frac{1}{N}\sum_{\mathbf{q,q^\prime,q^\dprime}}& \Big(a^\dagger_{a,\mathbf{q}} \langle a_{a,\mathbf{q^\dprime}} a^\dagger_{b,\mathbf{-q-q^\prime+q^\dprime}} \rangle a^\dagger_{a,\mathbf{q^\prime}} \notag\\&+
    a^\dagger_{a,\mathbf{q}}\langle a^\dagger_{a,\mathbf{q^\prime}} a^\dagger_{b,\mathbf{-q-q^\prime+q^\dprime}}\rangle a_{a,\mathbf{q^\dprime}} 
    \notag \\& +
    a^\dagger_{a,\mathbf{q}} \langle a^\dagger_{a,\mathbf{q^\prime}} a_{a,\mathbf{q^\dprime}} \rangle a^\dagger_{b,\mathbf{-q-q^\prime+q^\dprime}}
    \notag\\& + 
    a^\dagger_{a,\mathbf{q^\prime}} \langle a^\dagger_{a,\mathbf{q}} a^\dagger_{b,\mathbf{-q-q^\prime+q^\dprime}} \rangle a_{a,\mathbf{q^\dprime}} 
    \notag \\& +
    a^\dagger_{a,\mathbf{q^\prime}} \langle a^\dagger_{a,\mathbf{q}} a_{a,\mathbf{q^\dprime}} \rangle a^\dagger_{b,\mathbf{-q-q^\prime+q^\dprime}} 
    \notag\\& +
    a_{a\mathbf{q^\dprime}}\langle a^\dagger_{a,\mathbf{q}} a^\dagger_{a,\mathbf{q^\prime}} \rangle a^\dagger_{b,\mathbf{-q-q^\prime+q^\dprime}} \Big)e^{i(\mathbf{q+q^\prime-q^\dprime})\cdot\mathbf{r}_j} \label{eq:HP-decoupling}.
\end{align}
Only terms that are diagonal in $\mathbf{q}$ are retained for the thermal averages, which will collapse one of the $q$-summations. In addition, terms with thermal averages of the kind $\langle a^\dagger_{a,\mathbf{q^\prime}} a^\dagger_{a(b),\mathbf{q^\prime}} \rangle$ (two raising operators), will yield decoupled terms of the form 
$$\frac{1}{N}\sum_{\mathbf{q,q^\prime}} a_{a,\mathbf{q}} \langle a^\dagger_{a,\mathbf{q^\prime}} a^\dagger_{a(b),\mathbf{q^\prime}} \rangle a^\dagger_{b(a),\mathbf{q}\mp2\mathbf{q^\prime}}e^{-i (\mathbf{q}\mp2\mathbf{q^\prime)}\cdot\mathbf{r}_j}.$$
Upon enforcing translational symmetry of the lattice, such terms will vanish in the limit of $N\rightarrow\infty$ since only terms with $\mathbf{q}\mp2\mathbf{q^\prime}=\pm\mathbf{q}$ are consistent with translational symmetry. The mean field treatment of the cubic Hamiltonian \eqref{eq:H_3} yields contributions linear in bosonic operators and can be ignored in analogy with the linear terms.
All fourth order terms in Eq. \eqref{eq:H_4} can be decoupled similar to Eq. \eqref{eq:HP-decoupling} and terms involving thermal averages of the form $\langle a_{a,\mathbf{q^\prime}} a_{b,\mathbf{q^\prime}} \rangle$ vanish similar to factors containing two raising operators.  
For the remaining thermal averages we define
\begin{align}\label{eq:n_bare}
   n_{ab\mathbf{q}} &= \langle a^\dagger_{a,\mathbf{q}} a_{b,\mathbf{q}} \rangle ,
\end{align}
and
\begin{align}
    \langle a_{b,\mathbf{q}} a^\dagger_{a,\mathbf{q}} \rangle = \delta_{ab} + n_{ab\mathbf{q}},
\end{align}
where we used the commutation properties of $a_{ai}$ and $a^\dagger_{bj}$. Decoupling all the quartic bosonic terms, and expressing the thermal averages using Eq. \eqref{eq:n_bare}, Eq. \eqref{eq:H_4} can be simplified to yield Eq. \eqref{eq:H1_q} with components given by
\begin{align}
    &\textrm{A}^1_{ab\mathbf{q}}  = \frac{1}{N} \sum_\mathbf{q^\prime}\bigg( \mathbf{u}^T_a\textrm{J}^\prime_{ab\mathbf{q}} \mathbf{u}^*_b \frac{n_{aa\mathbf{q^\prime}} + n_{bb\mathbf{q^\prime}}}{4} \notag\\&+  \delta_{ab} \sum_{c} \frac{\mathbf{u}^T_c\textrm{J}^\prime_{ca\mathbf{q^\prime}}\mathbf{u}^*_a n_{ca\mathbf{q^\prime}} + \mathbf{u}^T_a\textrm{J}^\prime_{ac\mathbf{q^\prime}} \mathbf{u}^*_c (n_{ac\mathbf{q^\prime}}+\delta_{ac}/2) }{4}\bigg), \\
    &\textrm{B}^1_{ab\mathbf{q}} = \frac{1}{N} \sum_\mathbf{q^\prime} \bigg( \mathbf{u}^T_a\textrm{J}^\prime_{ab\mathbf{q}} \mathbf{u}_b \frac{n_{aa\mathbf{q^\prime}} + n_{bb\mathbf{q^\prime}}}{4} \notag\\&+  \delta_{ab} \sum_{c} \frac{\mathbf{u}^T_a\textrm{J}^\prime_{ac\mathbf{q^\prime}} \mathbf{u}_c n_{ac\mathbf{q^\prime}} + \mathbf{u}^T_c\textrm{J}^\prime_{ac\mathbf{q^\prime}} \mathbf{u}_a (n_{ac\mathbf{q^\prime}}+\delta_{ac})}{8}\bigg), \\
    &\textrm{C}^1_{ab\mathbf{q}} = \frac{1}{N}  \sum_\mathbf{q^\prime} \bigg(\mathbf{v}^T_a \textrm{J}^\prime_{ab\mathbf{q-q^\prime}}\mathbf{v}_b (n_{ba\mathbf{q^\prime}} + \frac{\delta_{ab}}{2}) \notag\\& +\delta_{ab} \sum_c n_{cc\mathbf{q^\prime}} \mathbf{v}^T_a \textrm{J}^\prime_{ac\mathbf{0}}\mathbf{v}_c\bigg).\label{eq:HP1-compL}
\end{align}
This renders the Hamiltonian quadratic in bosonic operators (given an initial guess for \eqref{eq:n_bare}) and it may be para-diagonalized to yield Eq. \eqref{eq:H-hp-final}. This in turn provides us with the transformation matrix $\textrm{T}_\mathbf{q}$ \eqref{eq:T-def} that allow us to write
\begin{align}
   n_{ab\mathbf{q}} &= \langle a^\dagger_{a,\mathbf{q}} a_{b,\mathbf{q}} \rangle \notag\\
   &= \sum_n [\textrm{T}_\mathbf{q}]_{a+N_a, n+N_a} [\textrm{T}^*_\mathbf{q}]_{n+N_a,b+N_a} \langle \alpha^\dagger_{n,\mathbf{q}} \alpha_{n,\mathbf{q}}\rangle\\
   &= \sum_n [\textrm{T}_\mathbf{q}]_{a+N_a, n+N_a} [\textrm{T}^*_\mathbf{q}]_{n+N_a,b+N_a} n_\textrm{B}(\omega_{n,\mathbf{q}}), \label{eq:n_abq}
\end{align}
where we used the fact that the thermal average of the normal bosonic modes may be written as the bose occupation of the magnon eigenmodes. This may be inserted back into Eq. \eqref{eq:H1_q}, which allows one to obtain a self-consistent solution iteratively.

\section{RPA}\label{app:Gfn}
The Hamiltonian in Eq. \eqref{eq:rot-heisenberg-model0} can be expressed as
\begin{align}\label{eq:HfullRPA}
    \mathcal{H} = -\frac{1}{2}\sum_{abij}\Big(& \mathbf{u}^{*T}_a \textrm{J}^{\prime}_{abij}\mathbf{u}^*_b \frac{S^{\dprime +}_{ai}S^{\dprime +}_{bj}}{4}+\mathbf{u}^{*T}_a \textrm{J}^{\prime}_{abij}\mathbf{u}_b \frac{S^{\dprime +}_{ai}S^{\dprime -}_{bj}}{4} \notag\\+&
    \mathbf{u}^{T}_a \textrm{J}^{\prime}_{abij}\mathbf{u}^*_b \frac{S^{\dprime -}_{ai}S^{\dprime +}_{bj}}{4}+\mathbf{u}^{T}_a \textrm{J}^{\prime}_{abij}\mathbf{u}_b \frac{S^{\dprime -}_{ai}S^{\dprime -}_{bj}}{4}\notag\\+&
    \mathbf{u}^{T}_a \textrm{J}^{\prime}_{abij}\mathbf{v}_b \frac{S^{\dprime +}_{ai}S^{\dprime z}_{bj}}{2} + \mathbf{u}^{*T}_a \textrm{J}^{\prime}_{abij}\mathbf{v}_b \frac{S^{\dprime -}_{ai}S^{\dprime z}_{bj}}{2} \notag\\+&
    \mathbf{v}^{T}_a \textrm{J}^{\prime}_{abij}\mathbf{u}_b \frac{S^{\dprime z}_{ai}S^{\dprime +}_{bj}}{2} + \mathbf{v}^{T}_a \textrm{J}^{\prime}_{abij}\mathbf{u}^*_b \frac{S^{\dprime z}_{ai}S^{\dprime -}_{bj}}{2} \notag\\+&
    \mathbf{v}^T_a \textrm{J}^{\prime}_{abij}\mathbf{v}_b S^{\dprime z}_{ai}S^{\dprime z}_{bj} \Big),
\end{align}
and the equations of motion for the Green's functions are written as
\begin{align}
    (\omega + i\epsilon) G^{+-}_{abij}(\omega) &= 2\langle S^{\dprime z}_a \rangle \delta_{ab} \delta_{ij}  + \langle \langle [S^{\dprime +}_{ai}, \mathcal{H}] ; S^{\dprime -}_{bj} \rangle\rangle_\omega,\\
    (\omega + i\epsilon) G^{--}_{abij}(\omega) &=  \langle \langle [S^{\dprime -}_{ai}, \mathcal{H}] ; S^{\dprime -}_{bj} \rangle\rangle_\omega.
\end{align}
Representing the commutator terms in the condensed notation of Eq. \eqref{eq:Gen-comm}, this can be expressed as in Eqs. \eqref{eq:eqofmotionGfn+}-\eqref{eq:eqofmotionGfn-}. Here, we calculate the commutators explicitly, which gives
\begin{align}
    [S^{\dprime +}_{ai}, \mathcal{H}] = -\frac{1}{2}\sum_{ck}\Big(& \frac{\mathbf{u}^{T}_a \textrm{J}^{\prime}_{acik}\mathbf{u}^*_c( S^{\dprime +}_{ck}S^{\dprime z}_{ai} +  S^{\dprime z}_{ai}S^{\dprime +}_{ck}) }{2} \notag\\+& \frac{\mathbf{u}^{T}_a \textrm{J}^{\prime}_{acik}\mathbf{u}_c( S^{\dprime z}_{ai}S^{\dprime -}_{ck}+S^{\dprime -}_{ck}S^{\dprime z}_{ai})}{2}
    \notag\\-& 
    \mathbf{v}^T_a \textrm{J}^{\prime}_{acik}\mathbf{v}_b (S^{\dprime +}_{ai}S^{\dprime z}_{ck} + S^{\dprime z}_{ck}S^{\dprime +}_{ai}) 
    \notag\\-&
    \mathbf{v}^T_a \textrm{J}^{\prime}_{acik}\mathbf{u}_c S^{\dprime +}_{ai}S^{\dprime +}_{ck}
    \notag\\-&
    \mathbf{v}^T_a \textrm{J}^{\prime}_{acik}\mathbf{u}^*_c S^{\dprime +}_{ai}S^{\dprime -}_{ck}
    \notag\\+&
    2\mathbf{u}^{*T}_a \textrm{J}^{\prime}_{acik}\mathbf{v}_c S^{\dprime z}_{ai}S^{\dprime z}_{ck}\Big), \label{eq:S+-comm}
    \\
    [S^{\dprime -}_{ai}, \mathcal{H}] =\frac{1}{2}\sum_{ck}\Big(& \frac{\mathbf{u}^{*T}_a \textrm{J}^{\prime}_{acik}\mathbf{u}_c (S^{\dprime z}_{ai}S^{\dprime -}_{ck} +  S^{\dprime -}_{ck}S^{\dprime z}_{ai})}{2} \notag\\+& \frac{\mathbf{u}^{*T}_a \textrm{J}^{\prime}_{acik}\mathbf{u}^*_c (S^{\dprime z}_{ai}S^{\dprime +}_{ck} + S^{\dprime +}_{ck}S^{\dprime z}_{ai})}{2} \notag\\-& \mathbf{v}^T_a \textrm{J}^{\prime}_{acik}\mathbf{v}_b (S^{\dprime -}_{ai}S^{\dprime z}_{ck} + S^{\dprime z}_{ck}S^{\dprime -}_{ai})
    \notag\\-&
    \mathbf{v}^T_a \textrm{J}^{\prime}_{acik}\mathbf{u}^*_c S^{\dprime -}_{ai}S^{\dprime -}_{ck}
    \notag\\-&
    \mathbf{v}^T_a \textrm{J}^{\prime}_{acik}\mathbf{u}_c S^{\dprime -}_{ai}S^{\dprime +}_{ck}
    \notag\\+&
    2\mathbf{u}^{T}_a \textrm{J}^{\prime}_{acik}\mathbf{v}_c S^{\dprime z}_{ai}S^{\dprime z}_{ck}
    \Big).\label{eq:S--comm}
\end{align}
The commutators above generates higher order Green's functions in the equations of motion which has to be approximated. For this, we apply RPA where the spin-fluctuations along the \textit{local z}-axis are assumed to be small and can be replaced by the ensemble average for every site. According to this, the Green's functions arising from the first three terms of the commutators in Eqs. \eqref{eq:S+-comm}-\eqref{eq:S--comm} can be decoupled as in Eq. \eqref{eq:RPA-decoupling}. The fourth and fifth terms give rise to Green's functions of the kind $\langle\langle S^{\dprime\pm}_{ai}S^{\dprime \pm}_{ck}; S^{\dprime -}_{bj}\rangle\rangle_\omega$ and $\langle\langle S^{\dprime\pm}_{ai}S^{\dprime \mp}_{ck}; S^{\dprime -}_{bj}\rangle\rangle_\omega$, which are negligible within the RPA framework (third order in transverse fluctuations), and are therefore discarded. The last term generates Green's functions of the kind $\langle\langle S^{\dprime z}_{ai}S^{\dprime z}_{ck}; S^{\prime -}_{bj}\rangle\rangle_\omega$ which also vanishes as both $S^{\dprime z}_{ck}$ and $S^{\dprime z}_{ai}$ are treated as constants. It is worth mentioning that these terms arise from the same terms in the Hamiltonian that gives rise to linear and cubic order bosonic terms in the HP method  (which also do not contribute to the dispersion). After applying RPA, the equations of motion in Eqs. \eqref{eq:eqofmotionGfn+}-\eqref{eq:eqofmotionGfn-} become
\begin{align}
    (\omega + i\epsilon) G^{+-}_{abij}(\omega) =& \ 2 \langle S^{\dprime z}_a \rangle \delta_{ab}\delta_{ij} \notag\\-&\frac{1}{2}\sum_{ck}\Big( \langle S^{\dprime z}_a \rangle \mathbf{u}^{T}_a \textrm{J}^{\prime}_{acik}\mathbf{u}_c G^{--}_{cbkj}(\omega) \notag \\&\qquad+\langle S^{\dprime z}_a \rangle \mathbf{u}^{T}_a \textrm{J}^{\prime}_{acik}\mathbf{u}^*_c  G^{+-}_{cbkj}(\omega) \notag\\&\qquad- 
    2 \langle S^{\dprime z}_c \rangle \mathbf{v}^T_a \textrm{J}^{\prime}_{abij}\mathbf{v}_b G^{+-}_{abij}(\omega) \Big),
\end{align}
\begin{align}
    (\omega + i\epsilon) G^{--}_{abij}(\omega) = \frac{1}{2}\sum_{ck}\Big(& \langle S^{\dprime z}_a \rangle \mathbf{u}^{*T}_a \textrm{J}^{\prime}_{acik}\mathbf{u}_c G^{--}_{cbkj}(\omega) \notag\\+& \langle S^{\dprime z}_a \rangle \mathbf{u}^{*T}_a \textrm{J}^{\prime}_{acik}\mathbf{u}^*_c G^{+-}_{cbkj}(\omega) - \notag \\& 2 \langle S^{\dprime z}_c \rangle \mathbf{v}^T_a \textrm{J}^{\prime}_{acik}\mathbf{v}_c G^{--}_{abij}(\omega) \Big).
\end{align}
This can be simplified to yield Eq. \eqref{eq:eqofmotionGfn} and the magnon energies can be obtained as the eigenvalues of the Hamiltonian \eqref{eq:H_rpa}.

\bibliography{references}

\begin{thebibliography}{42}%
\makeatletter
\providecommand \@ifxundefined [1]{%
 \@ifx{#1\undefined}
}%
\providecommand \@ifnum [1]{%
 \ifnum #1\expandafter \@firstoftwo
 \else \expandafter \@secondoftwo
 \fi
}%
\providecommand \@ifx [1]{%
 \ifx #1\expandafter \@firstoftwo
 \else \expandafter \@secondoftwo
 \fi
}%
\providecommand \natexlab [1]{#1}%
\providecommand \enquote  [1]{``#1''}%
\providecommand \bibnamefont  [1]{#1}%
\providecommand \bibfnamefont [1]{#1}%
\providecommand \citenamefont [1]{#1}%
\providecommand \href@noop [0]{\@secondoftwo}%
\providecommand \href [0]{\begingroup \@sanitize@url \@href}%
\providecommand \@href[1]{\@@startlink{#1}\@@href}%
\providecommand \@@href[1]{\endgroup#1\@@endlink}%
\providecommand \@sanitize@url [0]{\catcode `\\12\catcode `\$12\catcode `\&12\catcode `\#12\catcode `\^12\catcode `\_12\catcode `\%12\relax}%
\providecommand \@@startlink[1]{}%
\providecommand \@@endlink[0]{}%
\providecommand \url  [0]{\begingroup\@sanitize@url \@url }%
\providecommand \@url [1]{\endgroup\@href {#1}{\urlprefix }}%
\providecommand \urlprefix  [0]{URL }%
\providecommand \Eprint [0]{\href }%
\providecommand \doibase [0]{https://doi.org/}%
\providecommand \selectlanguage [0]{\@gobble}%
\providecommand \bibinfo  [0]{\@secondoftwo}%
\providecommand \bibfield  [0]{\@secondoftwo}%
\providecommand \translation [1]{[#1]}%
\providecommand \BibitemOpen [0]{}%
\providecommand \bibitemStop [0]{}%
\providecommand \bibitemNoStop [0]{.\EOS\space}%
\providecommand \EOS [0]{\spacefactor3000\relax}%
\providecommand \BibitemShut  [1]{\csname bibitem#1\endcsname}%
\let\auto@bib@innerbib\@empty
%</preamble>
\bibitem [{\citenamefont {Khomskii}(2009)}]{khomskii2009classifying}%
  \BibitemOpen
  \bibfield  {author} {\bibinfo {author} {\bibfnamefont {D.}~\bibnamefont {Khomskii}},\ }\bibfield  {title} {\bibinfo {title} {Classifying multiferroics: Mechanisms and effects},\ }\href@noop {} {\bibfield  {journal} {\bibinfo  {journal} {Physics}\ }\textbf {\bibinfo {volume} {2}},\ \bibinfo {pages} {20} (\bibinfo {year} {2009})}\BibitemShut {NoStop}%
\bibitem [{\citenamefont {\ifmmode~\check{S}\else \v{S}\fi{}mejkal}\ \emph {et~al.}(2022)\citenamefont {\ifmmode~\check{S}\else \v{S}\fi{}mejkal}, \citenamefont {Sinova},\ and\ \citenamefont {Jungwirth}}]{PhysRevX.12.040501}%
  \BibitemOpen
  \bibfield  {author} {\bibinfo {author} {\bibfnamefont {L.}~\bibnamefont {\ifmmode~\check{S}\else \v{S}\fi{}mejkal}}, \bibinfo {author} {\bibfnamefont {J.}~\bibnamefont {Sinova}},\ and\ \bibinfo {author} {\bibfnamefont {T.}~\bibnamefont {Jungwirth}},\ }\bibfield  {title} {\bibinfo {title} {Emerging research landscape of altermagnetism},\ }\href {https://doi.org/10.1103/PhysRevX.12.040501} {\bibfield  {journal} {\bibinfo  {journal} {Phys. Rev. X}\ }\textbf {\bibinfo {volume} {12}},\ \bibinfo {pages} {040501} (\bibinfo {year} {2022})}\BibitemShut {NoStop}%
\bibitem [{\citenamefont {Curie}(1894)}]{Curie1894}%
  \BibitemOpen
  \bibfield  {author} {\bibinfo {author} {\bibfnamefont {P.}~\bibnamefont {Curie}},\ }\bibfield  {title} {\bibinfo {title} {Sur la possibilité d'existence de la conductibilité magnétique et du magnétisme libre},\ }\href {https://doi.org/10.1051/jphystap:018940030041501} {\bibfield  {journal} {\bibinfo  {journal} {Journal de Physique}\ }\textbf {\bibinfo {volume} {3}},\ \bibinfo {pages} {415} (\bibinfo {year} {1894})}\BibitemShut {NoStop}%
\bibitem [{\citenamefont {Weiss}(1907)}]{Weiss1907}%
  \BibitemOpen
  \bibfield  {author} {\bibinfo {author} {\bibfnamefont {P.}~\bibnamefont {Weiss}},\ }\bibfield  {title} {\bibinfo {title} {L'hypothèse du champ moléculaire et la propriété ferromagnétique},\ }\href {https://doi.org/10.1051/jphystap:019070060066100} {\bibfield  {journal} {\bibinfo  {journal} {Journal de Physique}\ }\textbf {\bibinfo {volume} {6}},\ \bibinfo {pages} {661} (\bibinfo {year} {1907})}\BibitemShut {NoStop}%
\bibitem [{\citenamefont {Neel}(1936)}]{Neel1936}%
  \BibitemOpen
  \bibfield  {author} {\bibinfo {author} {\bibfnamefont {L.}~\bibnamefont {Neel}},\ }\bibfield  {title} {\bibinfo {title} {Propriétés magnétiques de l'état métallique et énergie d'interaction entre atomes magnétiques},\ }\href {https://doi.org/10.1051/anphys/193611050232} {\bibfield  {journal} {\bibinfo  {journal} {Annales de Physique}\ }\textbf {\bibinfo {volume} {11}},\ \bibinfo {pages} {232} (\bibinfo {year} {1936})}\BibitemShut {NoStop}%
\bibitem [{\citenamefont {Néel}(1971)}]{Neel1971}%
  \BibitemOpen
  \bibfield  {author} {\bibinfo {author} {\bibfnamefont {L.}~\bibnamefont {Néel}},\ }\bibfield  {title} {\bibinfo {title} {Magnetism and local molecular field},\ }\href {https://doi.org/10.1126/science.174.4013.985} {\bibfield  {journal} {\bibinfo  {journal} {Science}\ }\textbf {\bibinfo {volume} {174}},\ \bibinfo {pages} {985} (\bibinfo {year} {1971})}\BibitemShut {NoStop}%
\bibitem [{\citenamefont {Johnston}(2015)}]{Johnston2015}%
  \BibitemOpen
  \bibfield  {author} {\bibinfo {author} {\bibfnamefont {D.~C.}\ \bibnamefont {Johnston}},\ }\bibfield  {title} {\bibinfo {title} {Unified molecular field theory for collinear and noncollinear heisenberg antiferromagnets},\ }\href {https://doi.org/10.1103/PhysRevB.91.064427} {\bibfield  {journal} {\bibinfo  {journal} {Phys. Rev. B}\ }\textbf {\bibinfo {volume} {91}},\ \bibinfo {pages} {064427} (\bibinfo {year} {2015})}\BibitemShut {NoStop}%
\bibitem [{\citenamefont {Holstein}\ and\ \citenamefont {Primakoff}(1940)}]{hp1940}%
  \BibitemOpen
  \bibfield  {author} {\bibinfo {author} {\bibfnamefont {T.}~\bibnamefont {Holstein}}\ and\ \bibinfo {author} {\bibfnamefont {H.}~\bibnamefont {Primakoff}},\ }\bibfield  {title} {\bibinfo {title} {Field dependence of the intrinsic domain magnetization of a ferromagnet},\ }\href {https://doi.org/10.1103/PhysRev.58.1098} {\bibfield  {journal} {\bibinfo  {journal} {Phys. Rev.}\ }\textbf {\bibinfo {volume} {58}},\ \bibinfo {pages} {1098} (\bibinfo {year} {1940})}\BibitemShut {NoStop}%
\bibitem [{\citenamefont {Anderson}(1952)}]{Anderson1952}%
  \BibitemOpen
  \bibfield  {author} {\bibinfo {author} {\bibfnamefont {P.~W.}\ \bibnamefont {Anderson}},\ }\bibfield  {title} {\bibinfo {title} {An approximate quantum theory of the antiferromagnetic ground state},\ }\href {https://doi.org/10.1103/PhysRev.86.694} {\bibfield  {journal} {\bibinfo  {journal} {Phys. Rev.}\ }\textbf {\bibinfo {volume} {86}},\ \bibinfo {pages} {694} (\bibinfo {year} {1952})}\BibitemShut {NoStop}%
\bibitem [{\citenamefont {Marshall}\ and\ \citenamefont {Peierls}(1955)}]{Marshall1955}%
  \BibitemOpen
  \bibfield  {author} {\bibinfo {author} {\bibfnamefont {W.}~\bibnamefont {Marshall}}\ and\ \bibinfo {author} {\bibfnamefont {R.~E.}\ \bibnamefont {Peierls}},\ }\bibfield  {title} {\bibinfo {title} {The spin-wave theory of antiferromagnetism},\ }\href {https://doi.org/10.1098/rspa.1955.0201} {\bibfield  {journal} {\bibinfo  {journal} {Proceedings of the Royal Society of London. Series A. Mathematical and Physical Sciences}\ }\textbf {\bibinfo {volume} {232}},\ \bibinfo {pages} {69} (\bibinfo {year} {1955})}\BibitemShut {NoStop}%
\bibitem [{\citenamefont {Watabe}\ \emph {et~al.}(1995)\citenamefont {Watabe}, \citenamefont {Suzuki},\ and\ \citenamefont {Natsume}}]{Watabe1995}%
  \BibitemOpen
  \bibfield  {author} {\bibinfo {author} {\bibfnamefont {Y.}~\bibnamefont {Watabe}}, \bibinfo {author} {\bibfnamefont {T.}~\bibnamefont {Suzuki}},\ and\ \bibinfo {author} {\bibfnamefont {Y.}~\bibnamefont {Natsume}},\ }\bibfield  {title} {\bibinfo {title} {Theoretical study on quantum effects in triangular antiferromagnets with axial anisotropy using the numerically constructed bogoliubov transformation for magnons},\ }\href {https://doi.org/10.1103/PhysRevB.52.3400} {\bibfield  {journal} {\bibinfo  {journal} {Phys. Rev. B}\ }\textbf {\bibinfo {volume} {52}},\ \bibinfo {pages} {3400} (\bibinfo {year} {1995})}\BibitemShut {NoStop}%
\bibitem [{\citenamefont {Toth}\ and\ \citenamefont {Lake}(2015)}]{Toth2015}%
  \BibitemOpen
  \bibfield  {author} {\bibinfo {author} {\bibfnamefont {S.}~\bibnamefont {Toth}}\ and\ \bibinfo {author} {\bibfnamefont {B.}~\bibnamefont {Lake}},\ }\bibfield  {title} {\bibinfo {title} {Linear spin wave theory for single-q incommensurate magnetic structures},\ }\href {https://doi.org/10.1088/0953-8984/27/16/166002} {\bibfield  {journal} {\bibinfo  {journal} {Journal of Physics: Condensed Matter}\ }\textbf {\bibinfo {volume} {27}},\ \bibinfo {pages} {166002} (\bibinfo {year} {2015})}\BibitemShut {NoStop}%
\bibitem [{\citenamefont {Yosida}(2010)}]{yosida2010theory}%
  \BibitemOpen
  \bibfield  {author} {\bibinfo {author} {\bibfnamefont {K.}~\bibnamefont {Yosida}},\ }\href {https://books.google.dk/books?id=56CBcgAACAAJ} {\emph {\bibinfo {title} {Theory of Magnetism}}},\ Springer Series in Solid-State Sciences\ (\bibinfo  {publisher} {Springer Berlin Heidelberg},\ \bibinfo {year} {2010})\BibitemShut {NoStop}%
\bibitem [{\citenamefont {Rajeev~Pavizhakumari}\ \emph {et~al.}(2025)\citenamefont {Rajeev~Pavizhakumari}, \citenamefont {Skovhus},\ and\ \citenamefont {Olsen}}]{beyondrpa}%
  \BibitemOpen
  \bibfield  {author} {\bibinfo {author} {\bibfnamefont {V.}~\bibnamefont {Rajeev~Pavizhakumari}}, \bibinfo {author} {\bibfnamefont {T.}~\bibnamefont {Skovhus}},\ and\ \bibinfo {author} {\bibfnamefont {T.}~\bibnamefont {Olsen}},\ }\bibfield  {title} {\bibinfo {title} {Beyond the random phase approximation for calculating curie temperatures in ferromagnets: application to fe, ni, co and monolayer cri3},\ }\href {https://doi.org/10.1088/1361-648X/ada65c} {\bibfield  {journal} {\bibinfo  {journal} {Journal of Physics: Condensed Matter}\ }\textbf {\bibinfo {volume} {37}},\ \bibinfo {pages} {115806} (\bibinfo {year} {2025})}\BibitemShut {NoStop}%
\bibitem [{\citenamefont {{Bogolyubov}}\ and\ \citenamefont {{Tyablikov}}(1959)}]{tyablikov1959}%
  \BibitemOpen
  \bibfield  {author} {\bibinfo {author} {\bibfnamefont {N.~N.}\ \bibnamefont {{Bogolyubov}}}\ and\ \bibinfo {author} {\bibfnamefont {S.~V.}\ \bibnamefont {{Tyablikov}}},\ }\bibfield  {title} {\bibinfo {title} {{Retarded and Advanced Green Functions in Statistical Physics}},\ }\href@noop {} {\bibfield  {journal} {\bibinfo  {journal} {Soviet Physics Doklady}\ }\textbf {\bibinfo {volume} {4}},\ \bibinfo {pages} {589} (\bibinfo {year} {1959})}\BibitemShut {NoStop}%
\bibitem [{\citenamefont {Tahir-Kheli}\ and\ \citenamefont {ter Haar}(1962)}]{tahir1962}%
  \BibitemOpen
  \bibfield  {author} {\bibinfo {author} {\bibfnamefont {R.~A.}\ \bibnamefont {Tahir-Kheli}}\ and\ \bibinfo {author} {\bibfnamefont {D.}~\bibnamefont {ter Haar}},\ }\bibfield  {title} {\bibinfo {title} {Use of green functions in the theory of ferromagnetism. i. general discussion of the spin-$s$ case},\ }\href {https://doi.org/10.1103/PhysRev.127.88} {\bibfield  {journal} {\bibinfo  {journal} {Phys. Rev.}\ }\textbf {\bibinfo {volume} {127}},\ \bibinfo {pages} {88} (\bibinfo {year} {1962})}\BibitemShut {NoStop}%
\bibitem [{\citenamefont {Callen}(1963)}]{Callen1963}%
  \BibitemOpen
  \bibfield  {author} {\bibinfo {author} {\bibfnamefont {H.~B.}\ \bibnamefont {Callen}},\ }\bibfield  {title} {\bibinfo {title} {Green function theory of ferromagnetism},\ }\href {https://doi.org/10.1103/PhysRev.130.890} {\bibfield  {journal} {\bibinfo  {journal} {Phys. Rev.}\ }\textbf {\bibinfo {volume} {130}},\ \bibinfo {pages} {890} (\bibinfo {year} {1963})}\BibitemShut {NoStop}%
\bibitem [{\citenamefont {Lee}\ and\ \citenamefont {Liu}(1967)}]{Lee1967}%
  \BibitemOpen
  \bibfield  {author} {\bibinfo {author} {\bibfnamefont {K.~H.}\ \bibnamefont {Lee}}\ and\ \bibinfo {author} {\bibfnamefont {S.~H.}\ \bibnamefont {Liu}},\ }\bibfield  {title} {\bibinfo {title} {Green's-function method for antiferromagnetism},\ }\href {https://doi.org/10.1103/PhysRev.159.390} {\bibfield  {journal} {\bibinfo  {journal} {Phys. Rev.}\ }\textbf {\bibinfo {volume} {159}},\ \bibinfo {pages} {390} (\bibinfo {year} {1967})}\BibitemShut {NoStop}%
\bibitem [{\citenamefont {Turek}\ \emph {et~al.}(2003)\citenamefont {Turek}, \citenamefont {Kudrnovsk\'y}, \citenamefont {Divi\ifmmode~\check{s}\else \v{s}\fi{}}, \citenamefont {Franek}, \citenamefont {Bihlmayer},\ and\ \citenamefont {Bl\"ugel}}]{Turek2003}%
  \BibitemOpen
  \bibfield  {author} {\bibinfo {author} {\bibfnamefont {I.}~\bibnamefont {Turek}}, \bibinfo {author} {\bibfnamefont {J.}~\bibnamefont {Kudrnovsk\'y}}, \bibinfo {author} {\bibfnamefont {M.}~\bibnamefont {Divi\ifmmode~\check{s}\else \v{s}\fi{}}}, \bibinfo {author} {\bibfnamefont {P.}~\bibnamefont {Franek}}, \bibinfo {author} {\bibfnamefont {G.}~\bibnamefont {Bihlmayer}},\ and\ \bibinfo {author} {\bibfnamefont {S.}~\bibnamefont {Bl\"ugel}},\ }\bibfield  {title} {\bibinfo {title} {First-principles study of the electronic structure and exchange interactions in bcc europium},\ }\href {https://doi.org/10.1103/PhysRevB.68.224431} {\bibfield  {journal} {\bibinfo  {journal} {Phys. Rev. B}\ }\textbf {\bibinfo {volume} {68}},\ \bibinfo {pages} {224431} (\bibinfo {year} {2003})}\BibitemShut {NoStop}%
\bibitem [{\citenamefont {Bhowal}\ and\ \citenamefont {Spaldin}(2024)}]{MnF2-Altermagnet}%
  \BibitemOpen
  \bibfield  {author} {\bibinfo {author} {\bibfnamefont {S.}~\bibnamefont {Bhowal}}\ and\ \bibinfo {author} {\bibfnamefont {N.~A.}\ \bibnamefont {Spaldin}},\ }\bibfield  {title} {\bibinfo {title} {Ferroically ordered magnetic octupoles in $d$-wave altermagnets},\ }\href {https://doi.org/10.1103/PhysRevX.14.011019} {\bibfield  {journal} {\bibinfo  {journal} {Phys. Rev. X}\ }\textbf {\bibinfo {volume} {14}},\ \bibinfo {pages} {011019} (\bibinfo {year} {2024})}\BibitemShut {NoStop}%
\bibitem [{\citenamefont {Luttinger}\ and\ \citenamefont {Tisza}(1946)}]{Luttinger1946}%
  \BibitemOpen
  \bibfield  {author} {\bibinfo {author} {\bibfnamefont {J.~M.}\ \bibnamefont {Luttinger}}\ and\ \bibinfo {author} {\bibfnamefont {L.}~\bibnamefont {Tisza}},\ }\bibfield  {title} {\bibinfo {title} {Theory of dipole interaction in crystals},\ }\href {https://doi.org/10.1103/PhysRev.70.954} {\bibfield  {journal} {\bibinfo  {journal} {Phys. Rev.}\ }\textbf {\bibinfo {volume} {70}},\ \bibinfo {pages} {954} (\bibinfo {year} {1946})}\BibitemShut {NoStop}%
\bibitem [{\citenamefont {Schweflinghaus}\ \emph {et~al.}(2016)\citenamefont {Schweflinghaus}, \citenamefont {Zimmermann}, \citenamefont {Heide}, \citenamefont {Bihlmayer},\ and\ \citenamefont {Bl\"ugel}}]{PhysRevB.94.024403}%
  \BibitemOpen
  \bibfield  {author} {\bibinfo {author} {\bibfnamefont {B.}~\bibnamefont {Schweflinghaus}}, \bibinfo {author} {\bibfnamefont {B.}~\bibnamefont {Zimmermann}}, \bibinfo {author} {\bibfnamefont {M.}~\bibnamefont {Heide}}, \bibinfo {author} {\bibfnamefont {G.}~\bibnamefont {Bihlmayer}},\ and\ \bibinfo {author} {\bibfnamefont {S.}~\bibnamefont {Bl\"ugel}},\ }\bibfield  {title} {\bibinfo {title} {Role of dzyaloshinskii-moriya interaction for magnetism in transition-metal chains at pt step edges},\ }\href {https://doi.org/10.1103/PhysRevB.94.024403} {\bibfield  {journal} {\bibinfo  {journal} {Phys. Rev. B}\ }\textbf {\bibinfo {volume} {94}},\ \bibinfo {pages} {024403} (\bibinfo {year} {2016})}\BibitemShut {NoStop}%
\bibitem [{\citenamefont {Lyons}\ and\ \citenamefont {Kaplan}(1960)}]{Lyons1960}%
  \BibitemOpen
  \bibfield  {author} {\bibinfo {author} {\bibfnamefont {D.~H.}\ \bibnamefont {Lyons}}\ and\ \bibinfo {author} {\bibfnamefont {T.~A.}\ \bibnamefont {Kaplan}},\ }\bibfield  {title} {\bibinfo {title} {Method for determining ground-state spin configurations},\ }\href {https://doi.org/10.1103/PhysRev.120.1580} {\bibfield  {journal} {\bibinfo  {journal} {Phys. Rev.}\ }\textbf {\bibinfo {volume} {120}},\ \bibinfo {pages} {1580} (\bibinfo {year} {1960})}\BibitemShut {NoStop}%
\bibitem [{\citenamefont {Schmidt}\ and\ \citenamefont {Richter}(2022)}]{Schmidt2022}%
  \BibitemOpen
  \bibfield  {author} {\bibinfo {author} {\bibfnamefont {H.-J.}\ \bibnamefont {Schmidt}}\ and\ \bibinfo {author} {\bibfnamefont {J.}~\bibnamefont {Richter}},\ }\bibfield  {title} {\bibinfo {title} {Classical ground states of spin lattices},\ }\href {https://doi.org/10.1088/1751-8121/aca36d} {\bibfield  {journal} {\bibinfo  {journal} {Journal of Physics A: Mathematical and Theoretical}\ }\textbf {\bibinfo {volume} {55}},\ \bibinfo {pages} {465005} (\bibinfo {year} {2022})}\BibitemShut {NoStop}%
\bibitem [{\citenamefont {Colpa}(1978)}]{COLPA1978327}%
  \BibitemOpen
  \bibfield  {author} {\bibinfo {author} {\bibfnamefont {J.}~\bibnamefont {Colpa}},\ }\bibfield  {title} {\bibinfo {title} {Diagonalization of the quadratic boson hamiltonian},\ }\href {https://doi.org/https://doi.org/10.1016/0378-4371(78)90160-7} {\bibfield  {journal} {\bibinfo  {journal} {Physica A: Statistical Mechanics and its Applications}\ }\textbf {\bibinfo {volume} {93}},\ \bibinfo {pages} {327} (\bibinfo {year} {1978})}\BibitemShut {NoStop}%
\bibitem [{\citenamefont {Roth}(1958)}]{Roth1958-dist}%
  \BibitemOpen
  \bibfield  {author} {\bibinfo {author} {\bibfnamefont {W.~L.}\ \bibnamefont {Roth}},\ }\bibfield  {title} {\bibinfo {title} {Magnetic structures of mno, feo, coo, and nio},\ }\href {https://doi.org/10.1103/PhysRev.110.1333} {\bibfield  {journal} {\bibinfo  {journal} {Phys. Rev.}\ }\textbf {\bibinfo {volume} {110}},\ \bibinfo {pages} {1333} (\bibinfo {year} {1958})}\BibitemShut {NoStop}%
\bibitem [{\citenamefont {K\"odderitzsch}\ \emph {et~al.}(2002)\citenamefont {K\"odderitzsch}, \citenamefont {Hergert}, \citenamefont {Temmerman}, \citenamefont {Szotek}, \citenamefont {Ernst},\ and\ \citenamefont {Winter}}]{PhysRevB.66.064434}%
  \BibitemOpen
  \bibfield  {author} {\bibinfo {author} {\bibfnamefont {D.}~\bibnamefont {K\"odderitzsch}}, \bibinfo {author} {\bibfnamefont {W.}~\bibnamefont {Hergert}}, \bibinfo {author} {\bibfnamefont {W.~M.}\ \bibnamefont {Temmerman}}, \bibinfo {author} {\bibfnamefont {Z.}~\bibnamefont {Szotek}}, \bibinfo {author} {\bibfnamefont {A.}~\bibnamefont {Ernst}},\ and\ \bibinfo {author} {\bibfnamefont {H.}~\bibnamefont {Winter}},\ }\bibfield  {title} {\bibinfo {title} {Exchange interactions in nio and at the nio(100) surface},\ }\href {https://doi.org/10.1103/PhysRevB.66.064434} {\bibfield  {journal} {\bibinfo  {journal} {Phys. Rev. B}\ }\textbf {\bibinfo {volume} {66}},\ \bibinfo {pages} {064434} (\bibinfo {year} {2002})}\BibitemShut {NoStop}%
\bibitem [{\citenamefont {Kvashnin}\ \emph {et~al.}(2015)\citenamefont {Kvashnin}, \citenamefont {Gr\aa{}n\"as}, \citenamefont {Di~Marco}, \citenamefont {Katsnelson}, \citenamefont {Lichtenstein},\ and\ \citenamefont {Eriksson}}]{PhysRevB.91.125133}%
  \BibitemOpen
  \bibfield  {author} {\bibinfo {author} {\bibfnamefont {Y.~O.}\ \bibnamefont {Kvashnin}}, \bibinfo {author} {\bibfnamefont {O.}~\bibnamefont {Gr\aa{}n\"as}}, \bibinfo {author} {\bibfnamefont {I.}~\bibnamefont {Di~Marco}}, \bibinfo {author} {\bibfnamefont {M.~I.}\ \bibnamefont {Katsnelson}}, \bibinfo {author} {\bibfnamefont {A.~I.}\ \bibnamefont {Lichtenstein}},\ and\ \bibinfo {author} {\bibfnamefont {O.}~\bibnamefont {Eriksson}},\ }\bibfield  {title} {\bibinfo {title} {Exchange parameters of strongly correlated materials: Extraction from spin-polarized density functional theory plus dynamical mean-field theory},\ }\href {https://doi.org/10.1103/PhysRevB.91.125133} {\bibfield  {journal} {\bibinfo  {journal} {Phys. Rev. B}\ }\textbf {\bibinfo {volume} {91}},\ \bibinfo {pages} {125133} (\bibinfo {year} {2015})}\BibitemShut {NoStop}%
\bibitem [{\citenamefont {Olsen}(2017)}]{Olsen2017}%
  \BibitemOpen
  \bibfield  {author} {\bibinfo {author} {\bibfnamefont {T.}~\bibnamefont {Olsen}},\ }\bibfield  {title} {\bibinfo {title} {Assessing the performance of the random phase approximation for exchange and superexchange coupling constants in magnetic crystalline solids},\ }\href {https://doi.org/10.1103/PhysRevB.96.125143} {\bibfield  {journal} {\bibinfo  {journal} {Phys. Rev. B}\ }\textbf {\bibinfo {volume} {96}},\ \bibinfo {pages} {125143} (\bibinfo {year} {2017})}\BibitemShut {NoStop}%
\bibitem [{\citenamefont {Lines}\ and\ \citenamefont {Jones}(1965)}]{Lines1965MnO}%
  \BibitemOpen
  \bibfield  {author} {\bibinfo {author} {\bibfnamefont {M.~E.}\ \bibnamefont {Lines}}\ and\ \bibinfo {author} {\bibfnamefont {E.~D.}\ \bibnamefont {Jones}},\ }\bibfield  {title} {\bibinfo {title} {Antiferromagnetism in the face-centered cubic lattice. ii. magnetic properties of mno},\ }\href {https://doi.org/10.1103/PhysRev.139.A1313} {\bibfield  {journal} {\bibinfo  {journal} {Phys. Rev.}\ }\textbf {\bibinfo {volume} {139}},\ \bibinfo {pages} {A1313} (\bibinfo {year} {1965})}\BibitemShut {NoStop}%
\bibitem [{\citenamefont {Hutchings}\ and\ \citenamefont {Samuelsen}(1972)}]{Hutchings1972NiO}%
  \BibitemOpen
  \bibfield  {author} {\bibinfo {author} {\bibfnamefont {M.~T.}\ \bibnamefont {Hutchings}}\ and\ \bibinfo {author} {\bibfnamefont {E.~J.}\ \bibnamefont {Samuelsen}},\ }\bibfield  {title} {\bibinfo {title} {Measurement of spin-wave dispersion in nio by inelastic neutron scattering and its relation to magnetic properties},\ }\href {https://doi.org/10.1103/PhysRevB.6.3447} {\bibfield  {journal} {\bibinfo  {journal} {Phys. Rev. B}\ }\textbf {\bibinfo {volume} {6}},\ \bibinfo {pages} {3447} (\bibinfo {year} {1972})}\BibitemShut {NoStop}%
\bibitem [{\citenamefont {Alaei}\ and\ \citenamefont {Karimi}(2023)}]{Alaei2023}%
  \BibitemOpen
  \bibfield  {author} {\bibinfo {author} {\bibfnamefont {M.}~\bibnamefont {Alaei}}\ and\ \bibinfo {author} {\bibfnamefont {H.}~\bibnamefont {Karimi}},\ }\bibfield  {title} {\bibinfo {title} {A deep investigation of nio and mno through the first principle calculations and monte carlo simulations},\ }\href {https://doi.org/10.1088/2516-1075/acbff8} {\bibfield  {journal} {\bibinfo  {journal} {Electronic Structure}\ }\textbf {\bibinfo {volume} {5}},\ \bibinfo {pages} {025001} (\bibinfo {year} {2023})}\BibitemShut {NoStop}%
\bibitem [{\citenamefont {Stout}\ and\ \citenamefont {Reed}(1954)}]{doi:10.1021/ja01650a005}%
  \BibitemOpen
  \bibfield  {author} {\bibinfo {author} {\bibfnamefont {J.~W.}\ \bibnamefont {Stout}}\ and\ \bibinfo {author} {\bibfnamefont {S.~A.}\ \bibnamefont {Reed}},\ }\bibfield  {title} {\bibinfo {title} {The crystal structure of mnf2, fef2, cof2, nif2 and znf2},\ }\href {https://doi.org/10.1021/ja01650a005} {\bibfield  {journal} {\bibinfo  {journal} {Journal of the American Chemical Society}\ }\textbf {\bibinfo {volume} {76}},\ \bibinfo {pages} {5279} (\bibinfo {year} {1954})}\BibitemShut {NoStop}%
\bibitem [{\citenamefont {Stout}\ and\ \citenamefont {Matarrese}(1953)}]{RevModPhys.25.338}%
  \BibitemOpen
  \bibfield  {author} {\bibinfo {author} {\bibfnamefont {J.~W.}\ \bibnamefont {Stout}}\ and\ \bibinfo {author} {\bibfnamefont {L.~M.}\ \bibnamefont {Matarrese}},\ }\bibfield  {title} {\bibinfo {title} {Magnetic anisotropy of the iron-group fluorides},\ }\href {https://doi.org/10.1103/RevModPhys.25.338} {\bibfield  {journal} {\bibinfo  {journal} {Rev. Mod. Phys.}\ }\textbf {\bibinfo {volume} {25}},\ \bibinfo {pages} {338} (\bibinfo {year} {1953})}\BibitemShut {NoStop}%
\bibitem [{\citenamefont {Okazaki}\ \emph {et~al.}(1964)\citenamefont {Okazaki}, \citenamefont {Turberfield},\ and\ \citenamefont {Stevenson}}]{Okazaki1964}%
  \BibitemOpen
  \bibfield  {author} {\bibinfo {author} {\bibfnamefont {A.}~\bibnamefont {Okazaki}}, \bibinfo {author} {\bibfnamefont {K.}~\bibnamefont {Turberfield}},\ and\ \bibinfo {author} {\bibfnamefont {R.}~\bibnamefont {Stevenson}},\ }\bibfield  {title} {\bibinfo {title} {Neutron inelastic scattering measurements of antiferromagnetic excitations in mnf2 at 4.2°k and at temperatures up to the neel point},\ }\href {https://doi.org/https://doi.org/10.1016/0031-9163(64)90774-7} {\bibfield  {journal} {\bibinfo  {journal} {Physics Letters}\ }\textbf {\bibinfo {volume} {8}},\ \bibinfo {pages} {9} (\bibinfo {year} {1964})}\BibitemShut {NoStop}%
\bibitem [{\citenamefont {Durhuus}\ \emph {et~al.}(2023)\citenamefont {Durhuus}, \citenamefont {Skovhus},\ and\ \citenamefont {Olsen}}]{Durhuus_2023}%
  \BibitemOpen
  \bibfield  {author} {\bibinfo {author} {\bibfnamefont {F.~L.}\ \bibnamefont {Durhuus}}, \bibinfo {author} {\bibfnamefont {T.}~\bibnamefont {Skovhus}},\ and\ \bibinfo {author} {\bibfnamefont {T.}~\bibnamefont {Olsen}},\ }\bibfield  {title} {\bibinfo {title} {Plane wave implementation of the magnetic force theorem for magnetic exchange constants: application to bulk fe, co and ni},\ }\href {https://doi.org/10.1088/1361-648X/acab4b} {\bibfield  {journal} {\bibinfo  {journal} {Journal of Physics: Condensed Matter}\ }\textbf {\bibinfo {volume} {35}},\ \bibinfo {pages} {105802} (\bibinfo {year} {2023})}\BibitemShut {NoStop}%
\bibitem [{\citenamefont {Samuelsen}(1969)}]{Samuelsen1969swcor}%
  \BibitemOpen
  \bibfield  {author} {\bibinfo {author} {\bibfnamefont {E.}~\bibnamefont {Samuelsen}},\ }\bibfield  {title} {\bibinfo {title} {Spin waves in antiferromagnets with corundum structure},\ }\href {https://doi.org/https://doi.org/10.1016/0031-8914(69)90172-4} {\bibfield  {journal} {\bibinfo  {journal} {Physica}\ }\textbf {\bibinfo {volume} {43}},\ \bibinfo {pages} {353} (\bibinfo {year} {1969})}\BibitemShut {NoStop}%
\bibitem [{\citenamefont {Samuelsen}\ \emph {et~al.}(1969)\citenamefont {Samuelsen}, \citenamefont {Hutchings},\ and\ \citenamefont {Shirane}}]{Samuelsen1969Cr2O3}%
  \BibitemOpen
  \bibfield  {author} {\bibinfo {author} {\bibfnamefont {E.~J.}\ \bibnamefont {Samuelsen}}, \bibinfo {author} {\bibfnamefont {M.~T.}\ \bibnamefont {Hutchings}},\ and\ \bibinfo {author} {\bibfnamefont {G.}~\bibnamefont {Shirane}},\ }\bibfield  {title} {\bibinfo {title} {Inelastic neutron scattering investigation of spin waves and magnetic interactions in cr2o3},\ }\href {https://api.semanticscholar.org/CorpusID:119508376} {\bibfield  {journal} {\bibinfo  {journal} {Physica D: Nonlinear Phenomena}\ }\textbf {\bibinfo {volume} {48}},\ \bibinfo {pages} {13} (\bibinfo {year} {1969})}\BibitemShut {NoStop}%
\bibitem [{\citenamefont {Samuelsen}\ and\ \citenamefont {Shirane}(1970)}]{Samuelson1970Fe2O3}%
  \BibitemOpen
  \bibfield  {author} {\bibinfo {author} {\bibfnamefont {E.~J.}\ \bibnamefont {Samuelsen}}\ and\ \bibinfo {author} {\bibfnamefont {G.}~\bibnamefont {Shirane}},\ }\bibfield  {title} {\bibinfo {title} {Inelastic neutron scattering investigation of spin waves and magnetic interactions in $\alpha$-fe2o3},\ }\href {https://doi.org/https://doi.org/10.1002/pssb.19700420125} {\bibfield  {journal} {\bibinfo  {journal} {Physica status solidi (b)}\ }\textbf {\bibinfo {volume} {42}},\ \bibinfo {pages} {241} (\bibinfo {year} {1970})}\BibitemShut {NoStop}%
\bibitem [{\citenamefont {Loire}\ \emph {et~al.}(2011)\citenamefont {Loire}, \citenamefont {Simonet}, \citenamefont {Petit}, \citenamefont {Marty}, \citenamefont {Bordet}, \citenamefont {Lejay}, \citenamefont {Ollivier}, \citenamefont {Enderle}, \citenamefont {Steffens}, \citenamefont {Ressouche}, \citenamefont {Zorko},\ and\ \citenamefont {Ballou}}]{Loire2011}%
  \BibitemOpen
  \bibfield  {author} {\bibinfo {author} {\bibfnamefont {M.}~\bibnamefont {Loire}}, \bibinfo {author} {\bibfnamefont {V.}~\bibnamefont {Simonet}}, \bibinfo {author} {\bibfnamefont {S.}~\bibnamefont {Petit}}, \bibinfo {author} {\bibfnamefont {K.}~\bibnamefont {Marty}}, \bibinfo {author} {\bibfnamefont {P.}~\bibnamefont {Bordet}}, \bibinfo {author} {\bibfnamefont {P.}~\bibnamefont {Lejay}}, \bibinfo {author} {\bibfnamefont {J.}~\bibnamefont {Ollivier}}, \bibinfo {author} {\bibfnamefont {M.}~\bibnamefont {Enderle}}, \bibinfo {author} {\bibfnamefont {P.}~\bibnamefont {Steffens}}, \bibinfo {author} {\bibfnamefont {E.}~\bibnamefont {Ressouche}}, \bibinfo {author} {\bibfnamefont {A.}~\bibnamefont {Zorko}},\ and\ \bibinfo {author} {\bibfnamefont {R.}~\bibnamefont {Ballou}},\ }\bibfield  {title} {\bibinfo {title} {Parity-broken chiral spin dynamics in ${\mathrm{ba}}_{3}{\mathrm{nbfe}}_{3}{\mathrm{si}}_{2}{\mathrm{o}}_{14}$},\ }\href {https://doi.org/10.1103/PhysRevLett.106.207201} {\bibfield  {journal} {\bibinfo
  {journal} {Phys. Rev. Lett.}\ }\textbf {\bibinfo {volume} {106}},\ \bibinfo {pages} {207201} (\bibinfo {year} {2011})}\BibitemShut {NoStop}%
\bibitem [{\citenamefont {Olsen}(2024)}]{Olsen_2024}%
  \BibitemOpen
  \bibfield  {author} {\bibinfo {author} {\bibfnamefont {T.}~\bibnamefont {Olsen}},\ }\bibfield  {title} {\bibinfo {title} {Antiferromagnetism in two-dimensional materials: progress and computational challenges},\ }\href {https://doi.org/10.1088/2053-1583/ad4ef1} {\bibfield  {journal} {\bibinfo  {journal} {2D Materials}\ }\textbf {\bibinfo {volume} {11}},\ \bibinfo {pages} {033005} (\bibinfo {year} {2024})}\BibitemShut {NoStop}%
\bibitem [{\citenamefont {Litvin}(1974)}]{LITVIN1974205}%
  \BibitemOpen
  \bibfield  {author} {\bibinfo {author} {\bibfnamefont {D.}~\bibnamefont {Litvin}},\ }\bibfield  {title} {\bibinfo {title} {The luttinger-tisza method},\ }\href {https://doi.org/https://doi.org/10.1016/0031-8914(74)90257-2} {\bibfield  {journal} {\bibinfo  {journal} {Physica}\ }\textbf {\bibinfo {volume} {77}},\ \bibinfo {pages} {205} (\bibinfo {year} {1974})}\BibitemShut {NoStop}%
\end{thebibliography}%

\end{document}